\pdfoutput=1
\newif\ifExtendedVersion
\ExtendedVersiontrue

\ifExtendedVersion
\documentclass[sigconf,nonacm]{acmart}
\else
\documentclass[sigconf]{acmart}
\fi
\usepackage[utf8]{inputenc}
\usepackage[T1]{fontenc}

\newtheorem{finding}{Finding}
\newcommand{\snippet}[1]{\texttt{#1}}

\usepackage{subcaption}
\usepackage[inline]{enumitem}

\usepackage{tikz}
\usepackage{pgfplots}
\pgfplotsset{compat=1.17}
\usepackage{pgfplotstable}
\usepgfplotslibrary{statistics}
\usepgfplotslibrary{patchplots}
\definecolor{ColorCustomBlue}{rgb}{0,0,0.8}
\definecolor{ColorCustomRed}{rgb}{0.8,0,0}
\colorlet{ColorLegendPurple}{purple!90}
\colorlet{ColorLegendBrown}{brown!70}
\colorlet{ColorLegendBlue}{ColorCustomBlue!50}
\colorlet{ColorLegendRed}{ColorCustomRed!50}
\colorlet{ColorLegendGreen}{green!80!black}
\definecolor{ColorCustomGreen}{rgb}{0,0.3,0}
\pgfplotsset{
SmallBarPlot/.style={
    font=\footnotesize,
    ybar,
    width=\linewidth,
    ymin=0,
    xtick=data,
    xticklabel style={text width=0.8cm, align=center},
    xtick pos=left
},
BlueBigBars/.style={
    fill=blue!40, bar width=0.25
},
GreyBigBars/.style={
    fill=gray!40, bar width=0.25
},
RedBigBars/.style={
    fill=red!40, bar width=0.25
},
BlueBars/.style={
    fill=blue!40, bar width=0.2
},
DarkBlueBars/.style={
    fill=blue!90, bar width=0.2
},
RedBars/.style={
    fill=red!40, bar width=0.2
},
GreenBars/.style={
    fill=green!75!black, bar width=0.2
},
BrownBars/.style={
    fill=brown!70, bar width=0.2
},
PurpleBars/.style={
    fill=purple!90, bar width=0.2
},
OrangeBars/.style={
    fill=orange!40, bar width=0.2
},
PinkBars/.style={
    fill=pink!40, bar width=0.2
},
SmallBlueBars/.style={
    fill=blue!40, bar width=0.01
}
}

\pgfplotsset{select coords between index/.style 2 args={
    x filter/.code={
        \ifnum\coordindex<#1\fi
        \ifnum\coordindex>#2\fi
    }
}}

\pgfplotsset{
    ylabel right/.style={
        after end axis/.append code={
            \node [rotate=90, anchor=north] at (rel axis cs:1,0.5) {#1};
        }   
    }
}

\newcommand{\coloredsquare}[1]{\begin{tikzpicture}
\protect\tikz\protect\draw [color=#1, fill] (0,0) rectangle  (0.16,0.16);
\end{tikzpicture}\hspace{0.5pt}}

\usepackage[textsize=tiny, textwidth=1.5cm, colorinlistoftodos, disable]{todonotes}

\newcommand{\Aron}[1]{\todo[color=yellow!10, linecolor=yellow!75!black]{\textbf{Aron:} #1}}

\usepackage{hyperref}
\usepackage[capitalize]{cleveref}

\title{The Benefits of Vulnerability Discovery and Bug Bounty Programs: Case Studies of Chromium and Firefox}
\author{Soodeh Atefi}
\affiliation{\institution{University of Houston}
\country{USA}}
\author{Amutheezan Sivagnanam}
\affiliation{\institution{Pennsylvania State University}
\country{USA}}
\author{Afiya Ayman}
\affiliation{\institution{Pennsylvania State University}
\country{USA}}
\author{Jens Grossklags}
\affiliation{\institution{Technical University of Munich}
\country{Germany}}
\author{Aron Laszka}
\affiliation{\institution{Pennsylvania State University}
\country{USA}}

\copyrightyear{2023}
\acmYear{2023}
\setcopyright{acmlicensed}\acmConference[WWW '23]{Proceedings of the ACM Web Conference 2023}{April 30-May 4, 2023}{Austin, TX, USA}
\acmBooktitle{Proceedings of the ACM Web Conference 2023 (WWW '23), April 30-May 4, 2023, Austin, TX, USA}
\acmPrice{15.00}
\acmDOI{10.1145/3543507.3583352}
\acmISBN{978-1-4503-9416-1/23/04}


\begin{abstract}%
Recently, bug-bounty programs have gained popularity and become a significant
part of the security culture of many organizations. Bug-bounty programs enable
organizations to enhance their security posture by harnessing the diverse
expertise of crowds of external security experts (i.e., bug hunters). Nonetheless,
quantifying the benefits of bug-bounty programs remains elusive, which presents
a significant challenge for managing them. Previous studies focused on
measuring their benefits in terms of the number of vulnerabilities reported or
based on the properties of the reported vulnerabilities, such as severity or
exploitability. However, beyond these inherent properties, the value of a
report also depends on the probability that the vulnerability would be
discovered by a threat actor before an internal expert could discover and patch
it. In this paper, we present a data-driven study of the Chromium and Firefox
vulnerability-reward programs. First, we estimate the difficulty of discovering
a vulnerability using the probability of rediscovery as a novel metric. Our
findings show that vulnerability discovery and patching provide clear benefits
by making it difficult for threat actors to find vulnerabilities; however, we
also identify opportunities for improvement, such as incentivizing bug hunters
to focus more on development releases. Second, we compare the types of
vulnerabilities that are discovered internally vs. externally and those that
are exploited by threat actors. We observe significant differences between
vulnerabilities found by external bug hunters, internal security teams, and
external threat actors, which indicates that bug-bounty programs provide an
important benefit by complementing the expertise of internal teams, but also
that external hunters should be incentivized more to focus on the types of
vulnerabilities that are likely to be exploited by threat actors.%
\end{abstract}


\begin{document}

\ifExtendedVersion
\fancyhead[LE]{\footnotesize Published in the proceedings of the 2023 ACM Web Conference (WWW 2023).}
\fancyhead[LO]{\footnotesize The Benefits of Vulnerability Discovery and Bug Bounty Programs}
\fancyhead[RO]{\footnotesize Published in the proceedings of the 2023 ACM Web Conference (WWW 2023).}
\else
\fi

\maketitle
\setlength{\marginparwidth}{1.5cm}

\section{Introduction}
\label{sec:intro}

\ifExtendedVersion
Despite significant progress in software-engineering practices, the security of most software products and services remains imperfect in practice.
\fi
Traditionally, testing the security of software products and services was the responsibility of internal security teams and external penetration-testing teams. 
\ifExtendedVersion
However, these efforts are necessarily limited in their size and in the range of expertise applied.
This limitation puts defenders at a disadvantage compared to attackers since publicly-available products and services may be targeted by myriads of attackers, who possess diverse expertise (e.g., different attackers may be familiar with different techniques).

\fi
Spearheaded by Netscape as a forerunner in 1995~\cite{finifter2013empirical}, \emph{bug-bounty programs}---which are also known as \emph{vulnerability reward programs}---have emerged as a key element of many organizations' security culture~\cite{kuehn2014analyzing,mckinney2007vulnerability,zhao2015empirical}.
Bug-bounty programs are a form of crowdsourced vulnerability discovery, which
enables harnessing the diverse expertise of a large group of external bug hunters~\cite{finifter2013empirical}.
A program gives hackers the permission to test the security of a software product or service and to report vulnerabilities to the organization sponsoring the program \cite{laszka2018rules}.
By rewarding valid reports with bounties, the program incentivizes hackers to spend effort on searching for vulnerabilities and reporting them \cite{zhao2017devising, akgul2023bug}. %
In addition to enabling the sponsoring organization to fix security vulnerabilities before they could be exploited, a bug-bounty program also publicly signals the organization's commitment to continuously improving security.

However, \emph{quantifying the benefits of a bug-bounty program remains elusive}, which presents a significant challenge for managing them.
A number of prior research efforts have investigated bug-bounty programs (e.g., Finifter et al.~\cite{finifter2013empirical}, Zhao et al.~\cite{zhao2015empirical}, Laszka et al.~\cite{laszka2016banishing,laszka2018rules}, Maillart et al.~\cite{maillart2017given}, 
Luna et al.~\cite{luna2019productivity}, Elazari~\cite{elazari2019private}, Malladi and Subramanian~\cite{malladi2019bug}, and Walshe and Simpson~\cite{walshe2020empirical}).
However, a common limitation of previous studies is that they typically measure the value provided by a bug-bounty program in terms of the number of vulnerabilities reported or, in some cases, based on the inherent properties of the reported vulnerabilities, such as severity or exploitability.
As we discuss below, the number of reported vulnerabilities and their inherent properties alone cannot quantify security benefits since they ignore the \emph{likelihood of discovery}.

While some vulnerability reports provide immense value to organizations by enabling them to patch vulnerabilities before threat actors would exploit them, other reports might provide relatively low value.
First, \emph{some vulnerabilities might be discovered anyway by internal security experts} before any threat actors could exploit them.
Reports of such vulnerabilities provide low value since organizations could patch these vulnerabilities before exploitation without spending funds to reward external bug hunters.
Second, \emph{some vulnerabilities might never be discovered by threat actors}.
Patching such vulnerabilities is futile; in fact, it could even be considered harmful since patches can reveal the existence of vulnerabilities to threat actors~\cite{rescorla2005finding}.
Finally, even if some vulnerabilities are eventually discovered by threat actors, discovery might take so long that the software components become obsolete before the vulnerabilities could be exploited~\cite{Clark2014moving,roy2023survey}.
In contrast, other software projects may have relatively stable code bases over time, which also dominates the number of discovered vulnerabilities \cite{ozment2006milk}.
In light of these considerations, the value of a vulnerability report hinges not only on the inherent properties of the vulnerability, such as %
severity, but also on the \emph{probability that the reported vulnerability would be exploited by threat actors before another benign actor would report it}.

\paragraph{Research Questions}

\textbf{Benefits of vulnerability discovery (RQ1):} 
To study the issues mentioned above, we consider the probability of rediscovery, that is, the probability that a previously discovered vulnerabilities is independently rediscovered by another bug hunter.
The probability of rediscovery should be a key consideration for bug-bounty and vulnerability management since known vulnerabilities have a negative impact only if they are (re-)discovered by threat actors before they are patched (and before the patches are applied by users). 
\ifExtendedVersion
In fact, some prior works suggested that vulnerabilities should not be patched proactively because patching only brings them to the threat actors' attention. According to this view, proactive vulnerability patching and bug bounties would provide little value~\cite{rescorla2005finding}.
However, this proposition holds only if the probability of rediscovering a vulnerability is negligible. Schneier \cite{schneier2014should} conjectures, in contrast, 
\fi
that when one ``person finds a vulnerability, it is likely that another person soon will, or recently has, found the same vulnerability.''
Indeed, based on studying Microsoft security bulletins, Ozment finds that vulnerability rediscovery is non-negligible; but this result is based on a small sample (14 re-discovered vulnerabilities, constituting 7.69\% of all vulnerabilities listed in the bulletins)~\cite{ozment2005likelihood}.
In contrast, we characterize rediscovery probabilities based on thousands of vulnerability reports and thereby respond to Geer's call to conduct longitudinal research in this context \cite{geer2015good}.
\begin{itemize}[topsep=0pt]
    \item RQ1.1: Are vulnerabilities rediscovered? Are vulnerabilities more difficult to find, in terms of rediscovery probability, in stable releases than in development ones?
    \item RQ1.2: Are vulnerability discoveries and rediscoveries clustered in time, or is rediscovery a ``memory-less'' process?
\end{itemize}

\textbf{Benefits of bug bounties (RQ2):} 
If external bug hunters work similarly to internal security teams and discover similar vulnerabilities, then bug-bounty programs provide relatively low security benefits, and internalizing vulnerability-discovery efforts might be more efficient than sponsoring bug-bounty programs. However, theoretical work by Brady et al. suggests that there are efficiency benefits to testing software products in parallel by different teams that likely use different test cases and draw on different types of expertise \cite{brady1999murphy}. Supporting this view, Votipka et al.\ report key differences between internal security testers and external bug hunters based on a survey of 25 participants, focusing on how each group finds vulnerabilities,
how they develop their skills, and the challenges that they face~\cite{votipka2018hackers}.
In contrast, we focus on the actual vulnerabilities reported by these groups to facilitate the quantification of security benefits from the perspective of a sponsoring organization.
\begin{itemize}[topsep=0pt]
    \item RQ2: Do external bug hunters report different types of vulnerabilities than internal discoveries?
\end{itemize}

\textbf{Management of vulnerability discovery and bug bounties (RQ3):} 
The objective of both external and internal vulnerability discovery is to find and patch vulnerabilities that would be found by threat actors (since patching vulnerabilities that threat actors would not find provides a much lower security benefit).\footnote{Note that some bug hunters could be malicious; in this paper, we define bug hunter as someone who reports a vulnerability, thereby helping the program. At the same time, they might also try to exploit the vulnerability, which could be reported as the vulnerability being exploited in the wild. Since we focus on the benefits of vulnerability discovery, we study both activities strictly from the programs' perspective.} Hence, the benefits of running bug-bounty programs hinge on whether bug hunters find the same set of vulnerabilities that the threat actors would find. If there is a significant discrepancy, bug-bounty managers must try to steer bug hunters towards discovering the right types of vulnerabilities, e.g., using incentives.
\begin{itemize}
    \item RQ3.1: Do bug hunters report similar types of vulnerabilities than those that are being exploited by threat actors?
    \item RQ3.2: Which types of vulnerabilities are the most difficult to discover?
\end{itemize}

To answer these questions, we collect vulnerability data from the issue trackers of two major web browsers, Chromium (i.e., the open-source project that provides the vast majority of code for Google Chrome) and Firefox.
We combine these with other datasets and apply a thorough data cleaning process to reliably determine which reports are internal and which are external, %
which releases and components are impacted by each issue, which reports are duplicates (i.e., rediscoveries), %
which vulnerabilities were exploited, etc. 
Our cleaned datasets and our software implementation of the data collection and cleaning processes are publicly available~\cite{Atefi2023dataset}.

\paragraph{Organization} \Aron{TODO: probably cut from camera ready, keep in extended (decide once we know how much space we have)}The remainder of this paper is organized as follows.
\cref{sec:data} provides an overview of our data collection and cleaning processes.
\cref{sec:results} presents an in-depth analysis of the benefits of vulnerability discovery and bug-bounty programs.
\cref{sec:related} discusses related work on vulnerability discovery and bug bounties.
Finally, \cref{sec:concl} presents concluding remarks.

\section{Data Collection and Cleaning}
\label{sec:data}

We collect reports of security issues (i.e., vulnerabilities) from the issue trackers of %
Chromium and Firefox. 
An \emph{original report} of a vulnerability is a report that does not have \emph{duplicate} in its \emph{Status} field, which has typically---but not always---the earliest report date. %
A \emph{duplicate report}, identified by \emph{duplicate} in its \emph{Status} field, is a report of an issue that had already been reported. If the  duplicate and original reports were submitted by different bug hunters, then we consider the duplicate to be an independent \emph{rediscovery}. %

\Aron{we might be able to include the appendix as supplementary material, in which case we will need to change the phrasing (applies everywhere where we reference the arXiv version)}
\ifExtendedVersion
We describe the technical details of the data collection and cleaning processes in \cref{sec:data_collection,sec:data_cleaning}.
\fi

\subsection{Data Collection}

We collect the following five attributes for each vulnerability: impacted release channels (stable and/or development), security severity (critical, high, moderate, or low), weakness type represented as broad type of Common Weakness Enumeration (CWE), %
affected components, and programming languages (i.e., languages of files that were modified to patch the issue).
Note that for ease of exposition, we use the same names for severity levels and impacted releases for Chromium and Firefox; however, there is no one-to-one equivalence since there may be differences between the definitions of the two VRPs.
For a duplicate report, we use the attributes of the original report as 
the attributes of the duplicate. If an original report is missing some attributes, we use the attributes of its duplicates. %

\Aron{turn some of these footnotes to references to save space?}
\textbf{Chromium:}
We collect the details of all vulnerability reports from  September 2, 2008 -- September 13, 2022 from the Chromium issue tracker\footnote{\url{https://bugs.chromium.org/p/chromium/issues/}}  using Monorail API version 2\footnote{\url{https://chromium.googlesource.com/infra/infra/+/master/appengine/monorail/api/}}. For each report, the Chromium issue tracker lists affected components, impacted release channels, comments that include conversations among internal employees and external parties, as well as a history of changes (i.e., amendments) made to the report.

\textbf{Firefox:}
We collect data from two main resources, the Bugzilla Firefox bug tracker\footnote{\url{https://bugzilla.mozilla.org/home}} and the Mozilla website (Known Vulnerabilities\footnote{\url{https://www.mozilla.org/en-US/security/known-vulnerabilities/}} and Mozilla Foundation Security Advisories (MFSA)\footnote{\url{https://www.mozilla.org/en-US/security/advisories/}}). We collect security reports from January 24, 2012 -- September 15, 2022. 
The MFSA lists vulnerabilities for Mozilla products. We scrape advisories for Firefox to be able to identify reports that pertain to stable releases. 
We also collect the \emph{Reporter} field, which some pages in MFSA have, to identify external vs.\ internal reporters. %

\subsection{Data Cleaning}

\paragraph{Rediscovery and Duplicate Reports}

In both issue trackers, there is a \emph{Status} field that indicates whether a report is a duplicate of a previously reported vulnerability or an original report. %
In the Chromium issue tracker, for rediscoveries the \emph{Status} field is marked as \emph{Duplicate}. For each duplicate report, we follow the \emph{MergeInto} field to retrieve the referenced original report. If that is also a duplicate, we again follow the \emph{MergeInto} field of the referenced report until we find the original.
In the Firefox issue tracker, %
we can similarly determine whether a report is a duplicate based on the \emph{Status} field, and we can find the original report by following the references (recursively, if needed). %
Some vulnerabilities are reported multiple times by the same hunter; we remove these redundant reports and keep only the earliest report of each vulnerability for each hunter. Some vulnerabilities do not have accessible pages in Bugzilla. Since we cannot identify the earliest report for these vulnerabilities, we excluded them from our rediscovery analysis. Some Firefox reports are incomplete with respect to replication and patching. For some of these, Mozilla opened a new report of the vulnerability, which was then completed with respect to this information, and the first report was marked  as a duplicate. We also exclude these vulnerabilities from our analysis since they are not actual rediscoveries.

\paragraph{External vs. Internal Reports: Chromium}
The Chromium issue tracker contains reports of vulnerabilities either reported internally by Google or reported externally
by bug hunters. For each report, we use the reporter's email address to classify the report as either an \textit{internal} or an \textit{external report}. Note that we cannot always
determine the report's origin based solely on the email address. For each such email address, we manually check the associated activities, such as vulnerabilities reported and comments posted to determine the reporter's origin. We are able
to identify the email address of the actual external reporter for 98\% of the valid external reports.

\paragraph{External vs. Internal Reports: Firefox}
Vulnerabilities in Firefox VRP are reported either internally by Firefox team members or by external reporters. We use four steps to separate internal and external reports. First, we use the \emph{Whiteboard} and \emph{bug-flag} fields in the report information page. If a report has \emph{reporter-external} in \emph{Whiteboard} or \emph{sec-bounty} in \emph{bug-flag}, we generally consider the report to be  external; otherwise, we consider it to be internal. However, there are reports, which do not have the above keywords, but were reported by external bug hunters. To identify those reports, in the next step, we leverage a snow-balling technique (on the report comments) to identify around 650 reports that appear to be from internal team members of Mozilla on the first glance, but their actual reporters are external. In the third step, we consider reporters that appear to have both internal and external reports (around 50). We manually disambiguate these cases by reading comments and checking their public websites. In the last step, we leverage the \textit{Reporter} field in the MFSA by matching the reporters' profile names (from Bugzilla) with the names mentioned by the MFSA. By applying the above steps, we are able to %
distinguish internal and external reports in 97\% of the cases.

\paragraph{Stable vs. Development Release Channels}
Stable releases are widely used by end users, while development releases %
are typically used only by testers and bug hunters.
We use the term \emph{release channel} to refer to these different release versions. Note that we distinguish between reports that affect \emph{stable} releases %
and reports that affect only \emph{development} releases. %
In Chromium, there are  reports that affect both stable and development releases, which we exclude from our analysis of stable vs. development. For Firefox, we consider Bugzilla reports that are listed in the MFSA to be reports that affect stable releases.

\subsection{Other Data Sources}

Most vulnerabilities that have been fixed have links to the Google or Mozilla source-code repositories in their comments, which we use to identify the files that were changed to fix the vulnerability. For each vulnerability with a %
repository link, we collect the programming languages of the files that were changed. 
We also leverage CVEDetails\footnote{\url{https://www.cvedetails.com/}} and MITRE CWE\footnote{\url{https://cwe.mitre.org/}} to collect information regarding CVE IDs and weakness types (CWEs), when available. %

To identify exploited vulnerabilities, we first collect an initial set of exploited vulnerabilities from the Known Exploited Vulnerabilities Catalog of CISA\footnote{\url{https://www.cisa.gov/known-exploited-vulnerabilities-catalog}}. Then, we extend this set iteratively using a snowballing method by identifying terms in comments related to exploitation (e.g., \emph{exploited in the wild}) and gathering vulnerabilities whose comments include these terms.

\subsection{Summary of Datasets}

{\small\begin{table}[t]
\centering
\caption{Summary of Datasets}
\vspace{-1em}
\label{table:Summary}
\begin{tabular}{|c||r|r|}
\hline
\textbf{Security Severity} & \textbf{Chromium} & \textbf{Firefox}\\
\hline
Critical & 309 &  1,420\\ \hline
High &  8,616& 2,332\\ \hline
Moderate & 5,598 & 1,156 \\ \hline
Low &  2,720 & 605 \\ \hline
\hline
\textbf{Impacted Releases} & \textbf{Chromium} & \textbf{Firefox}\\\hline
Stable & 8,152  & 3,002\\ \hline
Development & 5,325 & 3,064  \\ \hline
\hline
\textbf{Reports} & \textbf{Chromium} & \textbf{Firefox}\\\hline
Duplicates& 3,905 & 1,262\\ \hline
Originals & 21,453 & 4,804 \\ \hline
\hline
\textbf{Reports' Origins} & \textbf{Chromium} & \textbf{Firefox}\\\hline
Externals & 12,221 & 1,837\\ \hline
Internals & 13,137 & 4,229\\ \hline
\end{tabular}
\label{exCounts}
\end{table}}

For Chromium, we collected  in total 25,358 valid reports of security issues. Of these, 
12,221 were externally reported, and 13,137 were internally reported. Among reports with information about impacted releases (13,477 reports in total), 8,152 reports pertain to stable releases, and 5,325 pertain to development one. Finally, 21,453 are original reports, and 3,905 are duplicate reports.
For Firefox, we collected in total 6,066 valid reports of security issues, of which 4,804 are original reports, and 1,262 are duplicates.
There are 3,002 reports of vulnerabilities which pertain to stable releases, and 3,064 reports that pertain to development releases. 1,837 reports were reported externally, and 4,229 were reported internally.
\cref{table:Summary} shows summary statistics of the two datasets.
\cref{table:stable_vs_development} shows the number of unique external bug hunters (note that 86 external reports were submitted anonymously for Firefox). %
For year-over-year temporal analysis, we provide annual data in \cref{additionaldata}.
We observe that most key metrics of interest are stable (e.g., fraction of duplicate reports is around 18.2\% for Chromium with 4.9\% standard deviation annually), which suggest that the reward programs' evolution over the past years does not significantly impact our findings.

{\small \begin{table}[t]
\centering
\caption{Comparison of Stable and Development Releases}
\label{table:stable_vs_development}
\vspace{-1em}
\begin{tabular}{|c||r|r|}
\hline
\textbf{Impacted Releases} & \textbf{Chromium} & \textbf{Firefox}\\\hline
\multicolumn{3}{|c|}{\textbf{Number of Unique External Reporters}} \\\hline
Stable & 1,297  & 413\\ \hline
Development & 198 & 285  \\ \hline\hline
\multicolumn{3}{|c|}{\textbf{Ratio of Rediscovered Vulnerabilities}} \\\hline
Stable & 8.63\%  & 9.27\%\\ \hline
Development & 4.26\% & 12.37\%  \\ \hline\hline
\multicolumn{3}{|c|}{\textbf{Mean Patching Time in Days}} \\\hline
Stable &  80.73 & 73.62\\ \hline
Development & 12.35 & 103.36\\ \hline
\end{tabular}
\end{table}}

\section{Results}
\label{sec:results}

\subsection{Benefits of Vulnerability Discovery and Bug Bounty Programs}

\subsubsection{Probability of Rediscovery (RQ 1.1)}

We begin our analysis by investigating whether vulnerabilities are more difficult to find in stable releases than in development ones.
To quantify this, we first study the probability that a vulnerability is rediscovered. %
\cref{table:stable_vs_development} shows  for each release channel the ratio of vulnerabilities that are rediscovered at least once. 
In Firefox, \emph{vulnerabilities that impact development releases are rediscovered more often than those that impact stable releases;} in Chromium, it is \emph{vice versa}.

Before drawing any conclusions about the difficulty of finding vulnerabilities, we must also consider the number of unique external reporters working on stable and development releases (see \cref{table:stable_vs_development}).
We find that in both Chromium and Firefox, \emph{there are considerably more bug hunters who report vulnerabilities in stable releases than in development ones}.
Combined with the rediscovery probabilities, this suggests that it is more difficult to find vulnerabilities in stable releases: although stable releases seem to attract significantly more attention, differences in rediscovery probabilities are less pronounced.

However, there is one more factor that can contextualize the difference in rediscovery probabilities: the amount of time required to patch a vulnerability. If it takes longer to patch a vulnerability, bug hunters have more time to rediscover it, which should lead to a higher rediscovery probability, ceteris paribus.
\cref{table:stable_vs_development} shows the average time between the first report of a vulnerability and the time when it was patched. %
We compute the time to patch $\Delta_\text{fix}$ %
as $\Delta_\text{fix} = T_\text{fix} - T_\text{earliest}$, where $T_\text{fix}$ is the date and time when the vulnerability was fixed and $T_\text{earliest}$ is when the vulnerability was first reported in the issue tracker. 
For Chromium, we observe that vulnerabilities in stable releases are patched much slower than in development releases, giving bug hunters significantly more time to rediscover them. %
For Firefox, the evidence is more nuanced. 
Here, we also observe a lower rediscovery probability for stable releases even though there is a larger workforce; however, hunters have to work with a slightly shorter average patching time window.

\begin{finding}
The rediscovery probabilities, number of bug hunters, and mean patching times in conjunction suggest that vulnerabilities are easier to find in development releases;
vulnerabilities that are easy to find are likely to be discovered and patched during development, demonstrating the benefits of vulnerability discovery. %
\end{finding}

\pgfplotstableread[col sep=comma,header=true]{data/Chrom/rediscovery/Chrom_re_fitted.csv}\Chromre
\pgfplotstableread[col sep=comma,header=true]{data/Chrom/rediscovery/Chrom_re_fitted_updated.csv}\ChromreFittingUpdate

\pgfplotstableread[col sep=comma,header=true]{data/Firefox/rediscovery/Firefox_re_fitted.csv}\Firefoxre
\pgfplotstableread[col sep=comma,header=true]{data/Firefox/rediscovery/Firefox_re_fitted_updated.csv}\FirefoxreFittingUpdate

\pgfplotstableread[col sep=comma,header=true]{data/Chrom/rediscovery/Chrom_re_weekly_stable_fitted.csv}\ChromreStable
\pgfplotstableread[col sep=comma,header=true]{data/Chrom/rediscovery/Chrom_re_weekly_stable_fitted_updated.csv}\ChromreStableFittingUpdate

\pgfplotstableread[col sep=comma,header=true]{data/Firefox/rediscovery/Firefox_re_weekly_stable_fitted.csv}\FirefoxreStable
\pgfplotstableread[col sep=comma,header=true]{data/Firefox/rediscovery/Firefox_re_weekly_stable_fitted_updated.csv}\FirefoxreStableFittingUpdate

\pgfplotstableread[col sep=comma,header=true]{data/Firefox/rediscovery/Firefox_re_weekly_notstable_fitted.csv}\FirefoxreNotStable
\pgfplotstableread[col sep=comma,header=true]{data/Firefox/rediscovery/Firefox_re_weekly_notstable_fitted_update.csv}\FirefoxreNotStableFittingUpdate

\pgfplotstableread[col sep=comma,header=true]{data/Chrom/rediscovery/Chrom_re_weekly_notstable_fitted.csv}\ChromreNotStable
\pgfplotstableread[col sep=comma,header=true]{data/Chrom/rediscovery/Chrom_re_weekly_notstable_fitted_updated.csv}\ChromreNotStableFittingUpdate

\begin{figure}[t]
\begin{subfigure}[b]{\linewidth}
     \begin{tikzpicture}[font=\small]
    \begin{axis}[xmin=1,xmax=100,
                 ymin=0, ymax=0.02,
                 xlabel=Number of Days $t$ from $T_\text{earliest}$, 
  ylabel=Probability {},
         width=\textwidth,
        height = 3cm,
        legend pos=north east,
        legend style={legend columns=-1,font=\tiny}
        ]

        \addplot [only marks, color = blue, mark = +]table [x=Days, y=Prob] {\ChromreFittingUpdate};
        \addlegendentry{C}  
  \addplot [thick, color = blue ]table [x=Days, y=FittedCurve] {\ChromreFittingUpdate};
      
               \addlegendentry{$0.01 \cdot t^{-0.70}$}
        \addplot [only marks, color = red, mark = x]table [x=Days, y=Prob] {\FirefoxreFittingUpdate};
      \addlegendentry{F}
   \addplot [thick, color = red ]table [x=Days, y=FittedCurve] {\FirefoxreFittingUpdate};
     \addlegendentry{$0.01\cdot t^{-0.61}$}

    \end{axis}
    \end{tikzpicture}
		\caption{Probability that a vulnerability is rediscovered on the $t$-th day after it is first reported ($\Pr\left[Re(t) \middle| t < \Delta_\text{fix}\right]$).}
	\label{subfig:rediscovery}
\end{subfigure}\vspace{0.25em}
\begin{subfigure}[b]{\linewidth}
     \begin{tikzpicture}[font=\small]
    \begin{axis}[xmin=1,xmax=13,ymin=0,ymax=0.035,xlabel=Number of Weeks $w$ from $T_\text{earliest}$,
  ylabel=Probability,
         width=\textwidth,
        height = 3cm, 
        legend pos=north east,
        legend style={legend columns=-1,font=\tiny}]

        \addplot [only marks, color = blue, mark = +]table [x=Weeks, y=Prob] {\ChromreStableFittingUpdate};
        \addlegendentry{C}  
  \addplot [thick, color = blue ]table [x=Weeks, y=FittedCurve] {\ChromreStableFittingUpdate};
             \addlegendentry{$0.02 \cdot w^{-0.60}$}

        \addplot [only marks, color = red, mark = x]table [x=Weeks, y=Prob] {\FirefoxreStableFittingUpdate};
      \addlegendentry{F}
   \addplot [thick, color = red ]table [x=Weeks, y=FittedCurve] {\FirefoxreStableFittingUpdate};
     \addlegendentry{$0.01 \cdot w ^ {-0.32}$}
    \end{axis}
    \end{tikzpicture}
		\caption{Probability that a vulnerability in a stable release is rediscovered in the $w$-th week after it is first reported ($\Pr\left[Re(w) \,\middle|\, w < \Delta_\text{fix}\right]$).}
	\label{subfig:re_Stable_Firefox_Chromium}
\end{subfigure}\vspace{0.25em}
\begin{subfigure}[b]{\linewidth}
     \begin{tikzpicture}[font=\small]
    \begin{axis}[xmin=1,xmax=13,ymin=0,ymax=0.06,xlabel=Number of Weeks $w$ from $T_\text{earliest}$,
  ylabel=Probability,
         width=\textwidth,
        height = 3cm, 
        legend pos=north east,
        legend style={legend columns=-1,font=\tiny}]

     \addplot [only marks, color = blue, mark = +]table [x=Weeks, y=Prob] {\ChromreNotStableFittingUpdate};
        \addlegendentry{C}  
  \addplot [thick, color = blue ]table [x=Weeks, y=FittedCurve] {\ChromreNotStableFittingUpdate};
             \addlegendentry{$0.02\cdot w^{-0.41}$}
             
        \addplot [only marks, color = red, mark = x]table [x=Weeks, y=Prob] {\FirefoxreNotStableFittingUpdate};
      \addlegendentry{F}
   \addplot [thick, color = red ]table [x=Weeks, y=FittedCurve] {\FirefoxreNotStableFittingUpdate};
     \addlegendentry{$0.04\cdot w^{-0.91}$}

    \end{axis}
    \end{tikzpicture}
		\caption{Probability that a vulnerability in a development release is rediscovered in the $w$-th week after it is first reported ($\Pr\left[Re(w) \,\middle|\, w < \Delta_\text{fix}\right]$).}
	\label{subfig:re_nonStable_Firefox_Chromium}
\end{subfigure}\vspace{0.1em}
\caption{Probability that a vulnerability is rediscovered a certain time after its first report, given that it has not been patched by that time. F and C denote Firefox and Chromium.} %
\label{fig:reStableVSDev}
\end{figure}
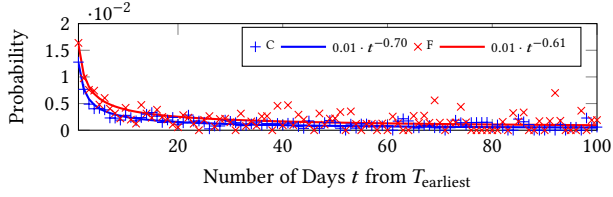
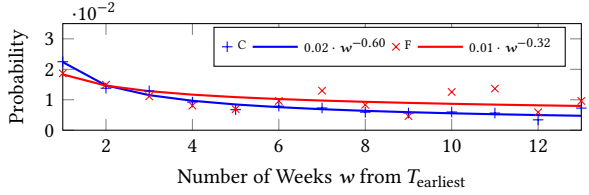
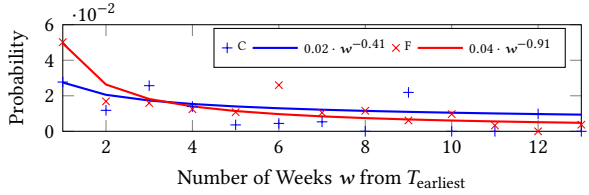
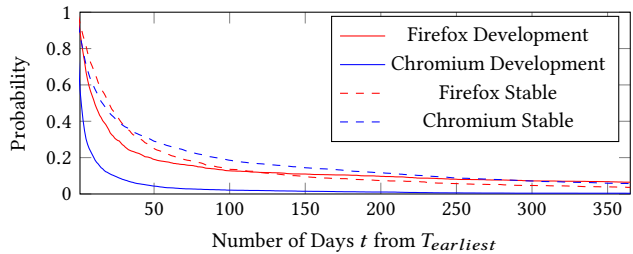
\begin{figure}[t]
\begin{tikzpicture}[font=\small]
	\begin{axis}[xmin=1,xmax=365,ymin=0,ymax=1,	xlabel=Number of Days $t$ from $T_{earliest}$,
        ylabel=Probability,
        width=0.5\textwidth,
        height = 4cm]
        \addplot [mark = none,red] table {Firefox_fix_development.dat};
         \addlegendentry{Firefox Development}
        \addplot[mark = none,blue] table {Chrom_fix_development.dat};
         \addlegendentry{Chromium Development}
      
        \addplot [dashed,mark = none,red] table {Firefox_fix_stable.dat};
         \addlegendentry{Firefox Stable}
        \addplot[dashed,mark = none,blue] table {Chrom_fix_stable.dat};
         \addlegendentry{Chromium Stable}
   
    	\end{axis}
\end{tikzpicture}
\vspace{-1em}
\caption{Probability that a vulnerability is not fixed in $t$ days after it is first reported ($\Pr\left[\Delta_{fix} > t\right]$) in \emph{development} (\emph{solid lines}) and \emph{stable} releases (\emph{dashed lines}).}
\label{fig:ProbNotFixedStableDevelopment}
\end{figure}
\begin{figure}
\begin{tikzpicture}[font=\small]
	\begin{axis}[xmin=1,xmax=365,ymin=0,ymax=1,	xlabel=Number of Days $t$ from $T_\text{earliest}$,
        ylabel=Probability,
        width=0.48\textwidth,
        height = 3.5cm]
        
        \addplot [mark = none,red] table {prob_not_fix_Firefox.dat};
         \addlegendentry{Firefox}
        \addplot[mark = none,blue] table {prob_not_fixed_Chrom.dat};
         \addlegendentry{Chromium}

      \addlegendentry{$\Pr\left[\Delta_{fix} > t\right]$}
    	\end{axis}

\end{tikzpicture}
 \vspace{-1em}
\caption{Probability that a vulnerability is not fixed in the $t$ days after it is first reported ($\Pr\left[\Delta_{fix} > t\right]$).}
\label{fig:ProbnotFixed}
\end{figure}
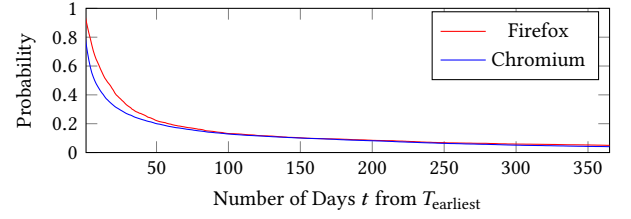\vspace{0.09em}
\pgfplotstableread[col sep=comma,header=true]{data/Chrom/rediscovery/Chrom_re_withoutcond_fitted.csv}\ChromreNoCond
\pgfplotstableread[col sep=comma,header=true]{data/Firefox/rediscovery/Firefox_re_withoutcond_fitted.csv}\FirefoxNoCond

\pgfplotstableread[col sep=comma,header=true]{data/Chrom/rediscovery/Chrom_re_withoutcond_fitted_update.csv}\ChromreNoCondFittingUpdate
\pgfplotstableread[col sep=comma,header=true]{data/Firefox/rediscovery/Firefox_re_withoutcond_fitted_updated.csv}\FirefoxNoCondFittingUpdate

\begin{figure}[!ht]
\begin{subfigure}[b]{\linewidth}
     \begin{tikzpicture}[font=\small]
    \begin{axis}[xmin=1,xmax=100,
                 ymin=0, ymax=0.02,
                 xlabel=Number of Days $t$ from $T_\text{earliest}$, 
  ylabel=Probability,
         width=\textwidth,
        height = 3.5cm,
         legend style={legend columns=2}
        ]

              \addplot [only marks, color = red, mark = x]table [x=Days, y=Prob] {\FirefoxNoCondFittingUpdate};
      \addlegendentry{Firefox}
   \addplot [thick, color = red ]table [x=Days, y=FittedCurve] {\FirefoxNoCondFittingUpdate};
     \addlegendentry{$0.01 \cdot t^{-0.82}$}
   
        \addplot [only marks, color = blue, mark = +]table [x=Days, y=Prob] {\ChromreNoCondFittingUpdate};
        \addlegendentry{Chromium}  
  \addplot [thick, color = blue ]table [x=Days, y=FittedCurve] {\ChromreNoCondFittingUpdate};
      
             \addlegendentry{$0.01 \cdot t^{-0.98}$}

    \end{axis}
    \end{tikzpicture}
\end{subfigure}
 \vspace{-2em}
\caption{Probability that a vulnerability is rediscovered on the $t$-th day after it is first reported ($\Pr[Re(t)]$).}
\label{fig:ReWithoutCond}
\end{figure}
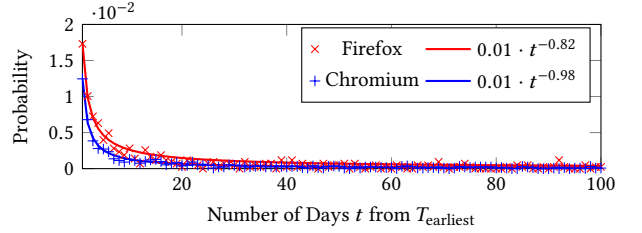

\subsubsection{Rediscovery Probability over Time (RQ 1.2)}

Since the probability of rediscovery alone cannot quantify the benefit of a vulnerability report, we contextualize the rediscovery probabilities of \emph{stable}, \emph{development}, and both releases together with the impact of patching.
For a duplicate report, we define the \emph{time until rediscovery} as %
$\Delta_{\text{rediscover}_d}=T_\text{open}-T_\text{earliest}$ (i.e., difference between the time of submitting the duplicate report and the submission time of the earliest report of the vulnerability).
We estimate the probability $\Pr\left[Re(t) \,\middle|\, t < \Delta_\text{fix}\right]$  that a vulnerability is rediscovered on the $t$-th day after it is first reported (this event is denoted $Re(t)$) given that the vulnerability has not been patched by day $t$ %
as follows:
\begin{equation}
\begin{split}
\left|\left\{ o_{d} \,|\,  d \in D, o_{d} \in O_\text{fix}, \Delta_{\text{rediscover}_{d}} = t\right\}\right|/\left|O_\text{fix}\right|
\end{split}
\end{equation}
The nominator is the number of original reports ($o_{d}$) that have not been fixed by day $t$ ($o_{d} \in O_\text{fix})$ and are rediscovered on the $t$-th day after they are first reported ($\Delta_{\text{rediscover}_{d}}=t$).
The denominator is the number of original reports that have not been fixed by day $t$.

Similarly, we also estimate probability $\Pr\left[Re(w) \,\middle|\, w < \Delta_\text{fix}\right]$ for the $w$-th  week %
as follows:
\begin{equation}
\begin{split}
\left|\left\{ o_{d} \,|\,  d \in D, o_{d} \in O_\text{fix}, 7 w -6 \leq \Delta_{\text{rediscover}_{d}} \leq 7 w\right\}\right|/\left|O_\text{fix}\right|
\end{split}
\end{equation}
where $O_\text{fix}\leftarrow\left\{o \in O \,|\, \Delta_{\text{fix}_{o}}> 7w \right\}$, i.e., not fixed by week $w$.

\cref{subfig:rediscovery} shows that rediscovery probabilities decrease over time in both Chromium and Firefox. We fit curves to identify and visualize important trends; building principled models that fit these trends is a major task, which we leave for future work.
When fitting curves, we weigh each probability value by its confidence, which we measure as the number of vulnerabilities based on which the value is estimated.
We also computed the probability that a vulnerability is not patched in $t$ days after it is first reported (see \cref{fig:ProbnotFixed}). This shows that 20\% of vulnerabilities are patched within 5 days of first being reported and that most vulnerabilities are patched quickly. %

Interestingly, even if we remove the condition ($t < \Delta_\text{fix}$) and consider the probability of rediscovery without the impact of patching, there is still a rapid decline in the first few days in both curves, i.e., Firefox and Chromium (see \cref{fig:ReWithoutCond}).
On the other hand, both curves have long tails later, which suggests a somewhat memory-less process of discovery (i.e., some vulnerabilities that are not discovered soon may remain hidden for long). %

\cref{subfig:re_Stable_Firefox_Chromium,subfig:re_nonStable_Firefox_Chromium} show the probability that a vulnerability is rediscovered in the $w$-th week after it is first reported (condition $w < \Delta_\text{fix}$) for vulnerabilities in stable and development releases, respectively. %
We observe that rediscovery probabilities are lower in stable releases than in development releases. 
In particular, this suggests that vulnerabilities in stable releases are more difficult to find and are non-trivial (see \cref{subfig:re_Stable_Firefox_Chromium}).
Further, the long tail suggests that vulnerabilities that have not yet been found may remain hidden for a long time, and discovery is mostly a memory-less process. 

However, there is a small peak in both curves in the first few days, which contradicts the memory-less property of rediscovery in stable releases, and suggests a clustering of rediscoveries. One hypothesis is that vendors pay more to external bug hunters for the discovery of vulnerabilities in stable releases relative to development releases. As a result, bug hunters may stockpile vulnerabilities in development releases to receive higher rewards by reporting vulnerabilities in stable releases, which increases the likelihood of duplicate reports. We tested this hypothesis by checking the reward policies of these two vendors. However, we did not find evidence for a higher reward policy for stable releases in either Chromium or Firefox. Instead, the reward policies are mostly based on vulnerability severity. Another hypothesis for the small peak in both curves (specifically for Chromium) is that vulnerabilities are more likely to be discovered shortly after they are introduced; exploring this hypothesis would require further technical analysis. %

\cref{subfig:re_nonStable_Firefox_Chromium} shows rapid decline in the first few days. This suggests that most vulnerabilities in development releases are rediscovered in the first few days after they are introduced.
\begin{finding}
Vulnerability discoveries are clustered in time, which suggests that there is a limited pool of easy-and-quick-to-discover vulnerabilities. Other vulnerabilities may remain hidden for long.
\end{finding}

\subsubsection{Internal Discoveries and External Bug Hunters (RQ2)}

\pgfplotstableread[col sep=comma,header=true]{data/Chrom/ie/Chrom_ie_Broad_type.csv}\ChromBroadsIe

\pgfplotstableread[col sep=comma,header=true]{data/Chrom/ie/Chrom_ie_CompGroup.csv}\ChromComponentsIe

\pgfplotstableread[col sep=comma,header=true]{data/Chrom/ie/Chrom_ie_Languages.csv}\ChromLangsIe

\pgfplotstableread[col sep=comma,header=true]{data/Firefox/ie/Firefox_ie_Broad_type.csv}\FirefoxBroadsIe
\pgfplotstableread[col sep=comma,header=true]{data/Firefox/ie/Firefox_ie_Component.csv}\FirefoxComponentIe
\pgfplotstableread[col sep=comma,header=true]{data/Firefox/ie/Firefox_ie_Language.csv}\FirefoxLangsIe

\begin{figure}

\begin{subfigure}[b]{\linewidth}
\null\hfill\begin{tikzpicture}[font=\footnotesize]
\null\vspace{1em}%

\begin{axis}[
    width=0.625\textwidth,
    height = 4.2cm, 
    xbar, 
    xmin=0,
    ylabel style={align=center},
    ylabel=Weakness Type,
    xlabel style={align=center},
    xlabel=Percentage of Vulnerabilities,
    ytick=data,
    ytick align=inside,
    ytick style={draw=none},
    legend pos=north east,
    legend columns=2,
    xticklabel=\pgfmathprintnumber{\tick}\,$\%$,
    xtick pos=upper,xticklabel pos=upper,
    yticklabels from table={\ChromBroadsIe}{Broad_type}
    ]
    \addplot [GreenBars, yshift=0.5pt] table[ y expr=\coordindex, x expr=\thisrow{internalPercentage}] {\ChromBroadsIe};
    \addplot [RedBars, yshift=-0.5pt] table[ y expr=\coordindex, x expr=\thisrow{externalPercentage}] {\ChromBroadsIe};
\end{axis}
\end{tikzpicture}
\null\vspace{-0.5em}
\caption{Chromium Vulnerabilities by Weakness Types}
\label{subfig:ChromBroadsIe}
\end{subfigure}
\null\vspace{0em}%
\begin{subfigure}[b]{\linewidth}
\null\hfill\begin{tikzpicture}[font=\footnotesize]
\begin{axis}[
    legend style={at={(0.485, 0.99)},anchor=north east},
    xbar, 
    xmin=0,
    width=0.625\textwidth,
    height = 4.2cm, 
    ylabel=Component,
    xtick={0, 20, 40},
    ytick=data,
    ytick align=inside,
    ytick style={draw=none},
    legend pos=north east,
    legend columns=2,
    xticklabel=\pgfmathprintnumber{\tick}\,$\%$,
    xtick pos=upper,xticklabel pos=upper,
    yticklabels from table={\ChromComponentsIe}{CompGroup},
    ]
    \addplot [GreenBars, yshift=0.5pt] table[ y expr=\coordindex, x expr=\thisrow{internalPercentage}] {\ChromComponentsIe};
    \addplot [RedBars, yshift=-0.5pt] table[ y expr=\coordindex, x expr=\thisrow{externalPercentage}] {\ChromComponentsIe};
\end{axis}
\end{tikzpicture}
\null\vspace{-0.5em}
\caption{Chromium Vulnerabilities by Components}
\label{subfig:ChromComponentIe}
\end{subfigure}
\null\vspace{0em}%
\begin{subfigure}[b]{\linewidth}
\null\hfill\begin{tikzpicture}[font=\footnotesize]
\begin{axis}[
    width=0.625\textwidth,
    height = 4.2cm, 
    xbar, 
    xmin=0, %
    ylabel style={align=center},
    ylabel=Weakness Type,
    xlabel style={align=center},
    ytick=data,
    legend pos=north east,
    legend columns=2,
    xticklabel=\pgfmathprintnumber{\tick}\,$\%$,
    xtick pos=upper,xticklabel pos=upper,
    ytick align=inside,
    ytick style={draw=none},
    yticklabels from table={\FirefoxBroadsIe}{Broad_type}
    ]
    \addplot [GreenBars, yshift=0.5pt] table[ y expr=\coordindex, x expr=\thisrow{internalPercentage}] {\FirefoxBroadsIe};
    \addplot [RedBars, yshift=-0.5pt] table[ y expr=\coordindex, x expr=\thisrow{externalPercentage}] {\FirefoxBroadsIe};
\end{axis}
\end{tikzpicture}
\null\vspace{-0.5em}
\caption{Firefox Vulnerabilities by Weakness Types}
\label{subfig:FirefoxBroadtypessIe}
\end{subfigure}
\null\vspace{1em}%
\begin{subfigure}[b]{\linewidth}
\null\hfill\begin{tikzpicture}[font=\footnotesize]
\begin{axis}[
    legend style={at={(0.485, 0.99)},anchor=north east},
    xbar, 
    xmin=0,
    width=0.625\textwidth,
    height = 4.2cm, 
    ylabel=Component,
    ytick=data,
    ytick align=inside,
    ytick style={draw=none},
    legend pos=north east,
    legend columns=2,
    xticklabel=\pgfmathprintnumber{\tick}\,$\%$,
    xtick pos=upper,xticklabel pos=upper,
    yticklabels from table={\FirefoxComponentIe}{Component},
    ]
    \addplot [GreenBars, yshift=0.5pt] table[ y expr=\coordindex, x expr=\thisrow{internalPercentage}] {\FirefoxComponentIe};
    \addplot [RedBars, yshift=-0.5pt] table[ y expr=\coordindex, x expr=\thisrow{externalPercentage}] {\FirefoxComponentIe};
\end{axis}
\end{tikzpicture}
\null\vspace{-0.5em}
\caption{Firefox Vulnerabilities by Components}
\label{subfig:FirefoxComponentIe}
\end{subfigure}

\null\vspace{-3em}%
\caption{Comparison of internal  (\coloredsquare{ColorLegendGreen}) and external (\coloredsquare{ColorLegendRed})  security reports in Chromium and Firefox.%
} 
 \label{fig:ie}
\end{figure}
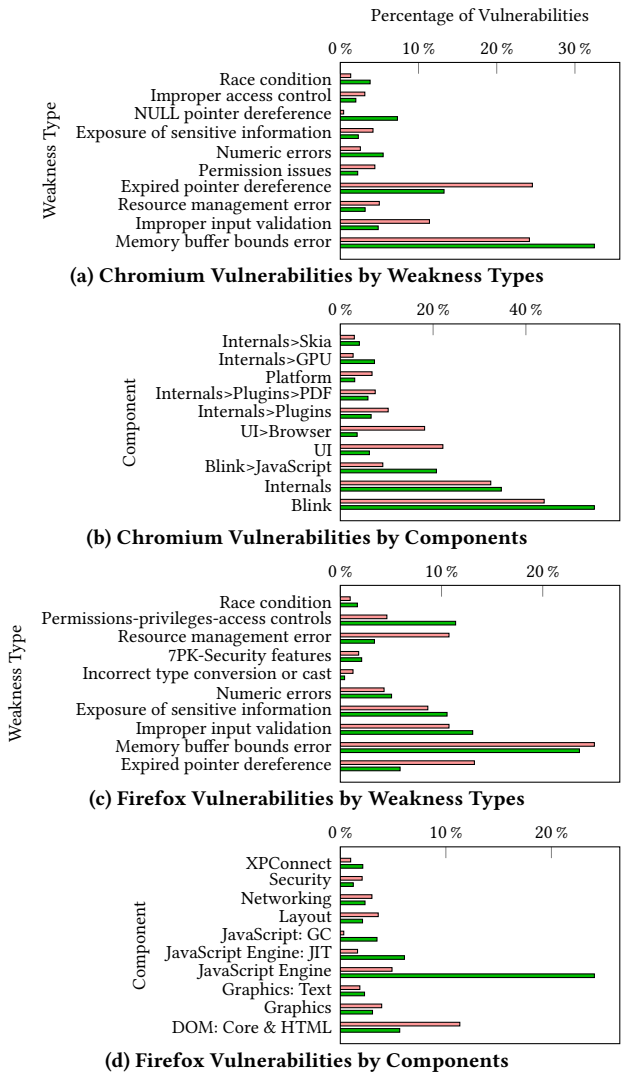

Next, we study the differences between reports of different origins (i.e., external versus internal) for Chromium and Firefox.
We provide a detailed comparison considering release channels in \cref{InExStable}.

\cref{fig:ie} shows the distributions of internal vs.\ external reports with respect to weakness types and impacted components. As for weakness types (see \cref{subfig:ChromBroadsIe}),
the most common type among internal Chromium reports is \emph{Memory buffer bound error} (32.5\%), while the most common type among external reports is \emph{Expired pointer dereference}.
In contrast, \emph{Memory buffer bound error} is the most common type among both internal and external Firefox reports (\cref{subfig:FirefoxBroadtypessIe}).
As for impacted components (see \cref{subfig:ChromComponentIe} and \cref{subfig:FirefoxComponentIe}), the \emph{Blink} component was most common among both internal and external Chromium reports; while in Firefox,  \emph{DOM: Core \& HTML} is the most common impacted component among external reports and \emph{JavaScript Engine: JIT} is the most common among internal~ones.

We also compared internal and external reports in terms of impacted release channels, severity, and programming languages. In Chromium, \emph{stable releases} are impacted by a higher percentage of reports than other releases, for both internal (50.1\%) and external (78.9\%) reports. In Firefox, we observe that 57.8\% of external reports pertain to \emph{stable releases}, while 54.1\% of internal reports relate to \emph{development releases}. As for severity, a high percentage of both internal and external reports have a \emph{high severity}. 
Further, external reports are more common than internal reports among vulnerabilities with \emph{critical severity}. We also find that vulnerabilities in \emph{C++} code are most frequently reported both internally and externally and in both Chromium and Firefox.

Based on Pearson's chi-squared test, %
external and internal reports follow significantly different distributions in terms of impacted release channels, severity level, weakness type, affected components, and programming languages for both Chromium and Firefox.

\begin{finding}
External bug hunters and internal security teams report different types of vulnerabilities, which indicates that bug-bounty programs do complement the expertise of internal teams. %
\end{finding}

\subsection{Management of Bug Bounty Programs}

\subsubsection{Vulnerabilities Reported and Exploited (RQ 3.1)}

\pgfplotstableread[col sep=comma]{data/Chrom/exploited/Chrom_ie_exploited_externalsComponent.csv}\ChromExComponent
\pgfplotstableread[col sep=comma]{data/Chrom/exploited/Chrom_ie_exploited_externalsBroad_type.csv}\ChromExBroad
\pgfplotstableread[col sep=comma]{data/Chrom/exploited/Chrom_ie_exploited_externalsLanguages.csv}\ChromExLanguages

\pgfplotstableread[col sep=comma]{data/Firefox/exploited/Firefox_exploited_other_ex_Component.csv}\FirefoxExComponent
\pgfplotstableread[col sep=comma]{data/Firefox/exploited/Firefox_exploited_other_ex_Broad_type.csv}\FirefoxExBroad
\pgfplotstableread[col sep=comma]{data/Firefox/exploited/Firefox_exploited_externals_Language.csv}\FirefoxExLanguages

\begin{figure}

\begin{subfigure}[b]{\linewidth}
\null\hfill\begin{tikzpicture}[font=\footnotesize]
\null\vspace{0em}%
\begin{axis}[
     width=0.625\textwidth,
    height = 3.5cm, 
    xbar, 
    xmin=0, %
    ylabel style={align=center},
    ylabel=Weakness Type,
    xlabel style={align=center},
    xlabel=Percentage of Vulnerabilities,
    ytick=data,
     ytick align=inside,
    ytick style={draw=none},
    legend pos=north east,
    legend columns=2,
    xticklabel=\pgfmathprintnumber{\tick}\,$\%$,
    xtick pos=upper,xticklabel pos=upper,
    yticklabels from table={\ChromExBroad}{Broad_type}
    ]
    \addplot [PurpleBars,yshift=0.5pt] table[ y expr=\coordindex, x expr=\thisrow{ExploitedPercentage}] {\ChromExBroad};
    \addplot [RedBigBars,yshift=-0.5pt] table[ y expr=\coordindex, x expr=\thisrow{OtherExPercentage}] {\ChromExBroad};
\end{axis}
\end{tikzpicture}
\null\vspace{-0.5em}
\caption{Chromium Vulnerabilities by Weakness Types}
\label{subfig:ChromBroadsEx}
\end{subfigure}
\null\vspace{2em}%
\begin{subfigure}[b]{\linewidth}
\null\hfill\begin{tikzpicture}[font=\footnotesize]
\begin{axis}[
    legend style={at={(0.485, 0.99)},anchor=north east},
    xbar, xmin=0,width=0.625\textwidth,
    height = 4.5cm, 
   xtick={0, 20, 40},
    ylabel=Component,
     ylabel style={align=center},
    ytick=data,
     ytick align=inside,
    ytick style={draw=none},
    legend pos=north east,
    legend columns=2,
    xticklabel=\pgfmathprintnumber{\tick}\,$\%$,
    xtick pos=upper,xticklabel pos=upper,
    yticklabels from table={\ChromExComponent}{CompGroup},
    ]
    \addplot [PurpleBars,yshift=0.5pt] table[ y expr=\coordindex, x expr=\thisrow{ExploitedPercentage}] {\ChromExComponent};
    \addplot [RedBigBars,yshift=-0.5pt] table[ y expr=\coordindex, x expr=\thisrow{OtherExPercentage}] {\ChromExComponent};
\end{axis}
\end{tikzpicture}
\null\vspace{-0.5em}
\caption{Chromium Vulnerabilities by Components}
\label{subfig:ChromComponentEx}
\end{subfigure}
\vspace{-3em}

\begin{subfigure}[b]{\linewidth}
\null\hfill\begin{tikzpicture}[font=\footnotesize]
\null\vspace{1em}%

\begin{axis}[
    width=0.625\textwidth,
    height = 3.5cm, 
    xbar, 
    xmin=0, %
    ylabel style={align=center},
    ylabel=Weakness Type,
    xlabel style={align=center},
     ytick align=inside,
    ytick style={draw=none},
    ytick=data,
    legend pos=north east,
    legend columns=2,
    xticklabel=\pgfmathprintnumber{\tick}\,$\%$,
    xtick pos=upper,xticklabel pos=upper,
    yticklabels from table={\FirefoxExBroad}{Broad_type}
    ]
    \addplot [PurpleBars,yshift=0.5pt] table[ y expr=\coordindex, x expr=\thisrow{ExploitedPercentage}] {\FirefoxExBroad};
    \addplot [RedBigBars,yshift=-0.5pt] table[ y expr=\coordindex, x expr=\thisrow{OtherExPercentage}] {\FirefoxExBroad};
\end{axis}
\end{tikzpicture}
\null\vspace{-0.5em}
\caption{Firefox Vulnerabilities by Weakness Types}
\label{subfig:FirefoxBroadsEx}
\end{subfigure}
\null\vspace{1em}%
\begin{subfigure}[b]{\linewidth}
\null\hfill\begin{tikzpicture}[font=\footnotesize]

\begin{axis}[
    legend style={at={(0.485, 0.99)},anchor=north east},
    xbar, xmin=0,width=0.625\textwidth,
    height = 4.5cm, 
    ylabel=Component,
    ytick=data,
     ytick align=inside,
    ytick style={draw=none},
    legend pos=north east,
    legend columns=2,
    xticklabel=\pgfmathprintnumber{\tick}\,$\%$,
    xtick pos=upper,xticklabel pos=upper,
    yticklabels from table={\FirefoxExComponent}{Component},
    ]
    \addplot [PurpleBars] table[ y expr=\coordindex, x expr=\thisrow{ExploitedPercentage}] {\FirefoxExComponent};
    \addplot [RedBigBars] table[ y expr=\coordindex, x expr=\thisrow{OtherExPercentage}] {\FirefoxExComponent};
\end{axis}
\end{tikzpicture}
\null\vspace{-0.5em}
\caption{Firefox Vulnerabilities by Components}
\label{subfig:FirefoxComponentEx}
\end{subfigure}

\null\vspace{-3em}%
\caption{Comparison of exploited vulnerabilities (\coloredsquare{ColorLegendPurple})  and 
external security reports (\coloredsquare{ColorLegendBrown}) in Chromium and Firefox based on weakness types and impacted components.}
\label{fig:ExploitedIssues_vs_AllExIssues}
\end{figure}
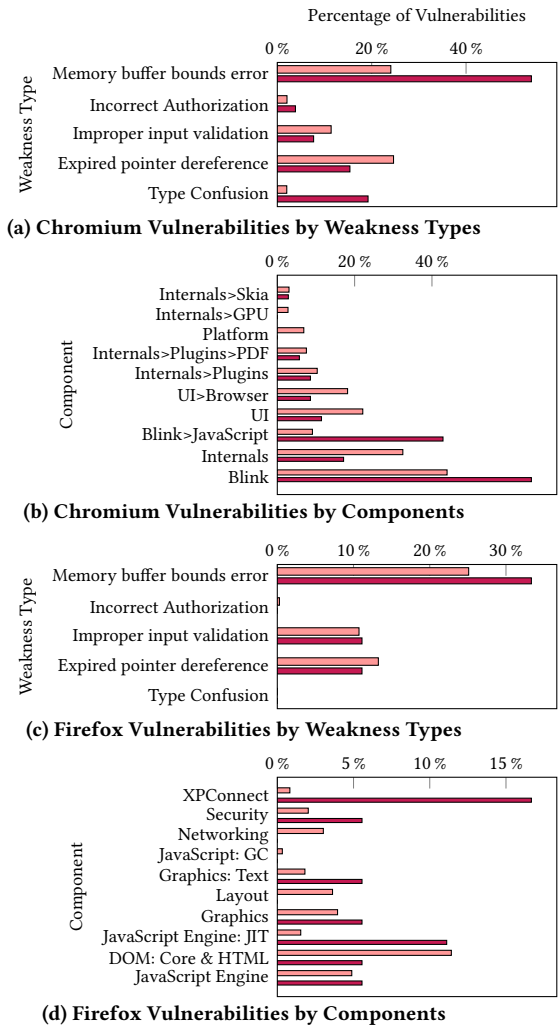

Finally, we study how many vulnerabilities have been discovered and exploited by malicious actors and what the differences are between these exploited vulnerabilities and other vulnerabilities.
Among the 25,358 (Chromium) and 6,066 (Firefox) valid vulnerability reports, we can identify 37 and 18 vulnerabilities that have been exploited in the wild for Chromium and Firefox, respectively. We compare these exploited vulnerabilities to those that are discovered by benevolent external reporters. We also compare these vulnerabilities with all other vulnerabilities (i.e., vulnerabilities that have not been exploited) based on release channels, severity, weakness type, components, and programming languages (see \cref{ExploitedOtherAppendix}). We perform chi-squared tests for all these comparisons as well. However, we acknowledge that the number of exploited vulnerabilities is limited; therefore, the results of our analysis might not be generalizable.

\paragraph{Comparison with Other Externally Reported Issues}
Since our focus is on bug-bounty programs, we study the differences and similarities between vulnerabilities that are exploited by threat actors and vulnerabilities that are reported by external bug hunters (\cref{fig:ExploitedIssues_vs_AllExIssues}). 
As for  release channels, all exploited Chromium vulnerabilities impact \emph{stable releases}, while only 78.9\% of external reports pertain to stable releases. In Firefox, 88.9\% of exploited vulnerabilities impact \emph{stable releases}, while only 57.5\% of external reports pertain to {stable} releases. With respect to severity,  71.4\% of exploited Chromium vulnerabilities are of \emph{high severity}, whereas only 45.3\% of external report have {high} severity. In Firefox, 62.5\% of exploited vulnerabilities have \emph{critical severity}, while only 25.8\% of external reports have \emph{critical} severity. Among both exploited vulnerabilities and external reports, vulnerabilities in \emph{C++} code were the most common. %

As for weakness types,
\emph{Memory buffer bound error} is the most commonly exploited type of vulnerability in both Chromium and Firefox (\cref{subfig:ChromBroadsEx,subfig:FirefoxBroadsEx}).
Fortunately, this weakness type is also very common among external reports in both Chromium and Firefox: it is the most common type in Firefox, and the second most common type in Chromium (just slightly behind the most common type, \emph{Expired pointer dereference}).
With respect to impacted components, 
the \emph{Blink} component of  Chromium is the most common among both exploited vulnerabilities and external reports (see \cref{subfig:ChromComponentEx}). %
In Firefox, the \emph{XPConnect} component is the most commonly impacted by exploited vulnerabilities; however, this component is relatively rare among external reports.

Our exploratory statistical tests show that for Chromium, exploited vulnerabilities and external reports follow significantly different distributions in terms of impacted release channels and security-severity levels; however, this does not hold for affected programming languages. %
Similarly, in Firefox, exploited vulnerabilities and external reports follow significantly different distributions in terms of impacted release channels and security severity. %

\begin{finding}
There are significant differences between the types of vulnerabilities that are reported by bug hunters and those that are exploited by threat actors in terms of impacted release channels, and security-severity levels, which suggests that bug bounties could be more effective if they incentivized bug hunters to shift their focus.
\end{finding}

\subsubsection{Difficulty of Discovery (RQ 3.2)}

\pgfplotstableread[col sep=comma,]{data/Chrom/rediscovery/re_ChromCompGroup.csv}\ReComponent
\pgfplotstableread[col sep=comma,]{data/Chrom/rediscovery/re_ChromBroad_type.csv}\ReBroads
\pgfplotstableread[col sep=comma,]{data/Chrom/rediscovery/re_ChromLanguages.csv}\ReLang

\pgfplotstableread[col sep=comma,]{data/Firefox/rediscovery/Firefox_remove_reopened_Component.csv}\ReComponentFirefox
\pgfplotstableread[col sep=comma,]{data/Firefox/rediscovery/Firefox_remove_reopened_Broad_type.csv}\ReBroadsFirefox
\pgfplotstableread[col sep=comma,]{data/Firefox/rediscovery/Firefox_remove_reopened_Languages.csv}\ReLangFirefox

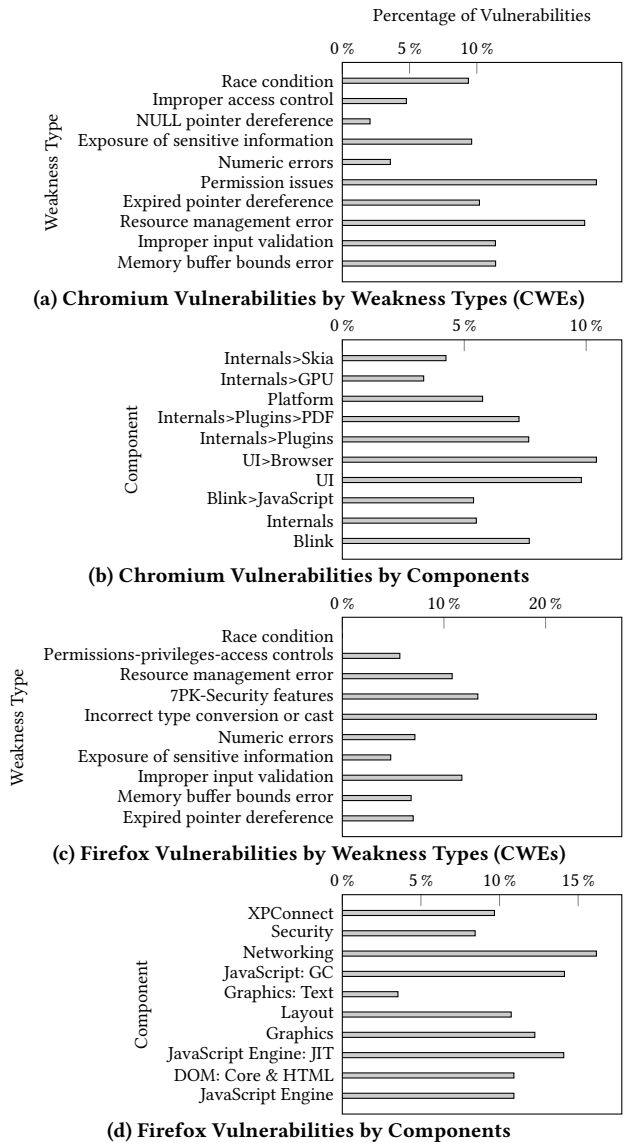
\begin{figure}
\begin{subfigure}[b]{\linewidth}
\null\hfill\begin{tikzpicture}[font=\footnotesize]
\null\vspace{1em}%
\begin{axis}[
    xbar,
    xmin = 0,
    width=0.625\textwidth,
    ylabel=Weakness Type,
    height = 4.5cm,
      xtick={0, 5,10},
      xlabel style={align=center},
   xlabel=Percentage of Vulnerabilities,
    ytick=data,
    ytick align=inside,
    ytick style={draw=none},
    legend pos=north east,
    xticklabel=\pgfmathprintnumber{\tick}\,$\%$,
    xtick pos=upper,xticklabel pos=upper,
    yticklabels from table={\ReBroads}{Broad_type},
    ]
     \addplot [GreyBigBars] table[ y expr=\coordindex, x expr=\thisrow{re_percentage}] {\ReBroads};
\end{axis}
\end{tikzpicture}
\null\vspace{-0.5em}
		\caption{Chromium Vulnerabilities by  Weakness Types (CWEs)}
	\label{subfig:reBroadsChrom}
\end{subfigure}
\null\vspace{-1.5em}%

\begin{subfigure}[b]{\linewidth}
\null\hfill\begin{tikzpicture}[font=\footnotesize]
\centering
\begin{axis}[
    xbar,
    xmin = 0,
     width=0.625\textwidth,
    ylabel=Component,
    height = 4.5cm,
    xtick={0, 5,10},
    ytick=data,
     ytick align=inside,
    ytick style={draw=none},
     legend pos=north east,
    xticklabel=\pgfmathprintnumber{\tick}\,$\%$,
    xtick pos=upper,xticklabel pos=upper,
    yticklabels from table={\ReComponent}{CompGroup},
    ]
     \addplot [GreyBigBars] table[ y expr=\coordindex, x expr=\thisrow{re_percentage}] {\ReComponent};
\end{axis}
\end{tikzpicture}
\null\vspace{-0.5em}
		\caption{Chromium Vulnerabilities by Components}
	\label{subfig:reComponentChrom}
\end{subfigure}
\null\vspace{-0.5em}%
\null\vspace{-1em}
\begin{subfigure}[b]{\linewidth}
\null\hfill\begin{tikzpicture}[font=\footnotesize]
\begin{axis}[
    xbar,
    xmin = 0,
     width=0.625\textwidth,
    ylabel=Weakness Type,
    height = 4.5cm,
    ytick=data,
      ytick align=inside,
    ytick style={draw=none},
    xticklabel=\pgfmathprintnumber{\tick}\,$\%$,
    xtick pos=upper,xticklabel pos=upper,
    yticklabels from table={\ReBroadsFirefox}{Broad_type},
    ]
     \addplot [GreyBigBars] table[ y expr=\coordindex, x expr=\thisrow{re_percentage}] {\ReBroadsFirefox};
\end{axis}
\end{tikzpicture}
\null\vspace{-0.5em}
		\caption{Firefox Vulnerabilities by Weakness Types (CWEs)}
	\label{subfig:reBroadtypesFirefox}
\end{subfigure}
\null\vspace{-1.5em}
\begin{subfigure}[b]{\linewidth}
\null\hfill\begin{tikzpicture}[font=\footnotesize]
\begin{axis}[
    xbar,
    xmin = 0,
    width=0.625\textwidth,
    ylabel=Component,
    height = 4.5cm,
    ytick=data,
      ytick align=inside,
    ytick style={draw=none},
    xticklabel=\pgfmathprintnumber{\tick}\,$\%$,
    xtick pos=upper,xticklabel pos=upper,
    yticklabels from table={\ReComponentFirefox}{Component},
    ]
     \addplot [GreyBigBars] table[ y expr=\coordindex, x expr=\thisrow{re_percentage}] {\ReComponentFirefox};
\end{axis}
\end{tikzpicture}
\null\vspace{-0.5em}
		\caption{Firefox Vulnerabilities by Components}
	\label{subfig:reComponentFirefox}
\end{subfigure}

\null\vspace{-1.5em}%
\caption{Fraction of vulnerabilities that are rediscovered at least once in Chromium and Firefox.}
\label{fig:RediscoveryRates}
\end{figure}

We estimate the probability of rediscovery as a function of the inherent properties of a vulnerability (i.e., security severity, weakness type, impacted components, and programming languages) to 
study whether different types of vulnerabilities are more or less difficult to rediscover (see \cref{fig:RediscoveryRates}). 
As for security severity, vulnerabilities with \emph{critical} and \emph{high} severity in Firefox and vulnerabilities with \emph{critical} severity in Chromium are rediscovered more than vulnerabilities with other severity levels. This can be partially explained by reward policies, which scale with the severity of the vulnerabilities. %
In Chromium, vulnerabilities with \emph{low} severity are rediscovered more than vulnerabilities with \emph{high} and \emph{moderate} severity levels. In Firefox, vulnerabilities with \emph{low} severity are rediscovered more than vulnerabilities with \emph{moderate} severity level. This may imply that vulnerabilities with \emph{low} severity are not only low-impact, but they are also shallow and easier to find. With respect to programming languages, vulnerabilities related to \emph{CSS} files in Chromium and \emph{Java} files in Firefox have  higher probabilities of rediscovery compared to vulnerabilities related to files in other languages.

As for weakness types in Chromium, vulnerabilities of the type \emph{Permission issues} are rediscovered more than vulnerabilities of other types (\cref{subfig:reBroadsChrom}). In Firefox, vulnerabilities of type \emph{Incorrect type conversion or cast} are more likely to be rediscovered than vulnerabilities of other types (\cref{subfig:reBroadtypesFirefox}). We also observe that vulnerabilities that impact the \emph{UI>Browser} component in Chromium and the \emph{Networking} component in Firefox are more likely to be rediscovered than vulnerabilities that impact other components (\cref{subfig:reComponentChrom,subfig:reComponentFirefox}). %
We also performed statistical tests %
between different types of vulnerabilities. The results show that there are significant differences between the rediscovery probabilities of different types of vulnerabilities. 

\begin{finding}
There are significant differences between the rediscovery probabilities of different types of vulnerabilities. %
Since vulnerabilities that are more severe than others receive higher rewards, and they are also rediscovered more often than other vulnerabilities, vendors could include other properties of vulnerabilities in their reward policy to incentivize external bug hunters.
\end{finding}

\section{Related Work}
\label{sec:related}

From a technical perspective, vulnerability discovery can be approached with a variety of static, dynamic, and concolic analysis methods as well as fuzzing~\cite{cui2022empirical,elder2022really}. Taking an analytical and empirical perspective, Massacci and Nguyen~\cite{massacci2014empirical} evaluated different mathematical vulnerability discovery models, which can be beneficial for vendors and users in terms of predicting vulnerability trends, adapting patching and update schedules, and allocating security investments. Prior works also investigated the empirical facets of vulnerability discovery in the context of bug bounty programs (e.g., \cite{elazari2019private,luna2019productivity,maillart2017given,walshe2022coordinated,zhao2015empirical,akgul2023bug,ding2019ethical}) and security bulletins (e.g., \cite{farhang19,farhang20}); however, research on rediscovery of vulnerabilities is sparse.
Ozment \cite{ozment2005likelihood} provided data on rediscovery frequency based on Microsoft's vulnerability bulletins and concluded that rediscovery is not negligible and should be explicitly considered in discovery models.
Finifter et al.~\cite{finifter2013empirical} studied VRPs; in one part of their study, they estimated average rediscovery rates of 4.6\% and similar rates for Chromium and Firefox, respectively. Both studies had to rely on very small datasets, but can serve as key motivators for our work. Herr et al. \cite{herr2017taking} estimated that vulnerability rediscovery occurs more often than previously reported (1\% to 9\%) in the literature (e.g., \cite{moussouris2015wolves}) and discuss patterns of rediscovery over time. Our work relies on a considerably more sizable dataset, which allows us to consider inherent patterns of rediscovery such as impacted release channels or weakness types. As such, our work goes well beyond the mere estimation of rediscovery rates.

Complementary to our investigation of vulnerability discovery, Iannone et al.\ \cite{iannone2022secret} study how, when, and under which circumstances vulnerabilities are introduced into software by developers and how they are removed.
While Iannone et al.\ studied the life-cycle of vulnerabilities by analyzing source code, Alomar et al.\ \cite{alomar2020you} conducted 53 interviews with security practitioners in technical and managerial roles to study vulnerability discovery and management processes in the wild.
In contrast, Akgul et al.\ \cite{akgul2023bug}, Votipka et al.\ \cite{votipka2018hackers}, and Fulton et al.\ \cite{fulton2022vulnerability} conducted surveys and interviews with bug hunters.
Alexopoulos et al.\ \cite{alexopoulos2021vulnerability} also studied bug hunters, but instead of conducting interviews, they collected information about a large number of bug hunters from public sources.

\section{Conclusion}
\label{sec:concl}

Our analysis illustrates that it is more difficult to rediscover vulnerabilities in \emph{stable} releases than in \emph{development} releases, considering all aspects of the process, including the number of bug hunters and the time-to-patch. Further, vulnerability discoveries and rediscoveries tend to be clustered in time after the first discovery, but seem to exhibit a long tail afterwards. %
In addition, the rediscovery probabilities of different types of vulnerabilities vary considerably. Likewise, our analysis shows that external bug hunters and internal staff and tools report different types of vulnerabilities, indicating that bug-bounty programs leverage the diverse expertise of external hackers. Furthermore, we discuss initial evidence regarding the difference between vulnerabilities that are exploited by threat actors and those found by external bug hunters.

\paragraph{Suggestions for Improving Bug Bounties}
Bug-bounty programs may benefit from incentivizing external hunters to focus more on development releases since the temporal clustering in stable releases suggest that some vulnerabilities that are relatively easy to find are not discovered during development.
Similarly, programs may benefit from incentivizing hunters to focus more on the types of
vulnerabilities that are likely to be exploited by threat actors.
Our analysis offers another important facet for the management of bug-bounty programs. Conducting the work to identify a vulnerability and filing a comprehensive report is a time-consuming matter. However, duplicate reports are typically not rewarded. As such, our work may provide guidance regarding how to channel the attention of bug hunters to avoid collisions or which patch development or triage efforts to prioritize to avoid hacker frustration.

\begin{acks}
This material is based upon work supported by the National Science Foundation under Grant No.\ CNS-1850510. Any opinions, findings, and conclusions or recommendations expressed in this material are those of the author(s) and do not necessarily reflect the views of the National Science Foundation.
We thank the anonymous reviewers for their valuable feedback and suggestions.
\end{acks}

\bibliography{main.bib}
\bibliographystyle{ACM-Reference-Format.bst}

\newpage
\appendix

\section{Additional Data}
\label{additionaldata}

\begin{table}[h!]
\centering
\caption{Number of Reports Per Year Based on Severity}
\label{table:SeverityPerYear}
\vspace{-1em}
\resizebox{\columnwidth}{!}{
\begin{tabular}{|c||c|c|c|c||c|c|c|c|}
\hline
\multicolumn{1}{|c||}{\textbf{Year Opened}}&  \multicolumn{4}{c||}{\textbf{Chromium}}  & \multicolumn{4}{c|}{\textbf{Firefox}} \\

\hline
& \textbf{Critical} & \textbf{High} & \textbf{Moderate} & \textbf{Low} & \textbf{Critical} & \textbf{High} & \textbf{Moderate}& \textbf{Low} \\\hline
2008       &          &      &          & 3  &&&& \\\hline
2009       & 5        & 52   & 33       & 46  &&&&\\\hline
2010       & 20       & 285  & 98       & 140 &&&&\\\hline
2011       & 54       & 528  & 134      & 133 &&&&\\\hline
2012       & 17       & 729  & 278      & 159  & 391      & 123  & 94       & 29 \\\hline
2013       & 20       & 544  & 242      & 154 & 331      & 215  & 92       & 56 \\\hline
2014       & 12       & 661  & 496      & 157 & 217      & 226  & 103      & 55 \\\hline
2015       & 10       & 675  & 282      & 174 & 170      & 263  & 124      & 63 \\\hline
2016       & 17       & 574  & 560      & 208  & 146      & 287  & 125      & 102\\\hline
2017       & 25       & 855  & 726      & 325  & 106      & 427  & 164      & 82  \\\hline
2018       & 24       & 942  & 772      & 365  & 30       & 236  & 113      & 71 \\\hline
2019       & 45       & 859  & 807      & 355 & 17       & 235  & 112      & 45\\\hline
2020       & 21       & 732  & 481      & 216  & 9        & 175  & 107      & 40 \\\hline
2021       & 31       & 900  & 547      & 207   &     0     & 128  & 84       & 39 \\\hline
2022       & 8        & 280  & 142      & 78  & 3        & 15   & 32       & 17 \\\hline
\end{tabular}
}
\end{table}
\begin{table}[h!]
\centering
\caption{Number of Reports Per Year Based on Releases}
\label{table:ReleasesPerYear}
\vspace{-1em}
\resizebox{\columnwidth}{!}{
\begin{tabular}{|c||c|c||c|c|}
\hline
\multicolumn{1}{|c||}{\textbf{Year Opened}}&  \multicolumn{2}{c||}{\textbf{Chromium}}  & \multicolumn{2}{c|}{\textbf{Firefox}} \\
\hline
 & \textbf{\# of Stable Reports} & \textbf{\# of Development Reports}  & \textbf{\# of Stable Reports} & \textbf{\# of Development Reports}\\\hline
2008       & 1                 &         &&       \\\hline
2009       & 95                & 1         &&     \\\hline
2010       & 361               &        0     &&   \\\hline
2011       & 524               & 27            && \\\hline
2012       & 716               & 103          & 302               & 396     \\\hline
2013       & 551               & 96            & 248               & 514      \\\hline
2014       & 776               & 338           & 270               & 413      \\\hline
2015       & 651               & 401         & 334               & 362        \\\hline
2016       & 637               & 538         & 376               & 331      \\\hline
2017       & 941               & 665       & 515               & 322      \\\hline
2018       & 875               & 794       & 281               & 217        \\\hline
2019       & 860               & 1066      & 233               & 209         \\\hline
2020       & 685               & 512      & 212               & 147        \\\hline
2021       & 435               & 624       & 177               & 117           \\\hline
2022       & 44                & 160         & 45                & 29       \\\hline
\end{tabular}
}
\end{table}

\begin{table}[h!]
\centering
\caption{Number of Original Vs. Duplicate Reports Per Year}
\label{table:DupOrgPerYear}
\vspace{-1em}
\resizebox{\columnwidth}{!}{
\begin{tabular}{|c||c|c||c|c|}
\hline
\multicolumn{1}{|c||}{\textbf{Year Opened}}&  \multicolumn{2}{c||}{\textbf{Chromium}}  & \multicolumn{2}{c|}{\textbf{Firefox}} \\
\hline
 & \textbf{\# of Duplicate Reports} & \textbf{\# of Original Reports} & \textbf{\# of Duplicate Reports} & \textbf{\# of Original Reports} \\\hline
2008       &                & 51         &&    \\\hline
2009       & 17             & 212          &&  \\\hline
2010       & 121            & 770            &&\\\hline
2011       & 231            & 991           && \\\hline
2012       & 330            & 1278    & 157            & 541           \\\hline
2013       & 218            & 1253        & 154            & 608        \\\hline
2014       & 275            & 1509         & 148            & 535      \\\hline
2015       & 320            & 1243          & 147            & 549     \\\hline
2016       & 310            & 1642        & 153            & 554     \\\hline
2017       & 458            & 2156         & 241            & 596    \\\hline
2018       & 423            & 2292        & 92             & 406     \\\hline
2019       & 407            & 2742       & 82             & 360        \\\hline
2020       & 313            & 2064    & 43             & 316              \\\hline
2021       & 373            & 2422      & 26             & 268         \\\hline
2022       & 109            & 828      & 3              & 71            \\
\hline
\end{tabular}
}
\end{table}

\begin{table}[h!]
\centering
\caption{Number of Reports Per Year Based on The Origins}
\label{table:ExInPerYear}
\vspace{-1em}
\resizebox{\columnwidth}{!}{
\begin{tabular}{|c||c|c||c|c|}
\hline
\multicolumn{1}{|c||}{\textbf{Year Opened}}&  \multicolumn{2}{c||}{\textbf{Chromium}}  & \multicolumn{2}{c|}{\textbf{Firefox}} \\
\hline
& \textbf{\# of Internal Reports} & \textbf{\# of External Reports} &\textbf{\# of Internal Reports} & \textbf{\# of External Reports} \\\hline
2008       & 23            & 28       &&     \\\hline
2009       & 146           & 83         &&   \\\hline
2010       & 455           & 436         &&  \\\hline
2011       & 632           & 590          && \\\hline
2012       & 867           & 741        & 502           & 196         \\\hline
2013       & 833           & 638      & 549           & 213         \\\hline
2014       & 1023          & 761      & 510           & 173       \\\hline
2015       & 762           & 801       & 460           & 236       \\\hline
2016       & 915           & 1037    & 461           & 246        \\\hline
2017       & 1590          & 1024  & 555           & 282            \\\hline
2018       & 1551          & 1164     & 329           & 169         \\\hline
2019       & 1908          & 1241      & 311           & 131      \\\hline
2020       & 1039          & 1338   & 275           & 84           \\\hline
2021       & 1130          & 1665     & 206           & 88          \\\hline
2022       & 263           & 674        & 57            & 17         \\\hline
\end{tabular}
}
\end{table}

\cref{table:SeverityPerYear,table:ReleasesPerYear,table:DupOrgPerYear,table:ExInPerYear} show annual data.

\begin{table}[h!]
\centering
\caption{Mean Patching Time in Days}
\label{table:AvgPatchingOtherFeatures}
\vspace{-1em}
\resizebox{\columnwidth}{!}{
\begin{tabular}{|c|r||c|r|}
\hline
\textbf{Chromium Security Severity} &\textbf{Days} & \textbf{Firefox Security Severity} &\textbf{Days}\\
\hline
Critical &28.46 & Critical& 23.56 \\ \hline
High &  38.25 &High& 55.26\\ \hline
Moderate & 47.14 &Moderate& 133.24\\ \hline
Low & 114.09 &Low&183.38\\ \hline
\hline
\textbf{Chromium Weakness Types} & \textbf{Days} &\textbf{Firefox Weakness Types} &\textbf{Days}\\
\hline
Race condition &53.27& Race condition&65.90 \\ \hline
Expired pointer dereference &30.17& Expired pointer dereference& 39.78\\ \hline
Memory buffer bounds error & 49.99&Memory buffer bounds error & 42.39\\ \hline
Improper input validation & 74.31 &Improper input validation&100.70\\ \hline
Exposure of sensitive information& 79.57&Exposure of sensitive information&107.61\\ \hline
Numeric errors& 49.44 & Numeric errors& 25.45\\ \hline
Permission issues&102.96&Incorrect type conversion or cast&24.75\\ \hline
Null pointer dereference&94.66&7PK\-Security features&143.80\\ \hline
Improper access control&60.67&Permissions-privileges-access controls&91.82\\ \hline
Resource management error&28.64&Resource management error&49.55\\ \hline
\hline
\textbf{Chromium Component} & \textbf{Days} &\textbf{Firefox Component}& \textbf{Days}\\\hline
Internals>Skia &  29.32 & XPConnect&120.77\\ \hline
Internals>GPU & 25.48 &Security &164.56\\ \hline
Platform & 90.85 & Networking&72.86\\ \hline
Internals>Plugins>PDF & 46.11&Layout &119.37\\ \hline
Internals>Plugins & 56.29 &JavaScript: GC &48.91\\ \hline
UI>Browser & 88.26 &JavaScript Engine: JIT&35.70\\ \hline
UI & 82.32 &JavaScript Engine&52.88\\ \hline
Blink>JavaScript& 13.27 & Graphics: Text&56.55\\ \hline
Internals& 48.20& Graphics&80.97\\ \hline
Blink & 51.58 & DOM: Core \& HTML&51.11\\ \hline
\hline
\textbf{Chromium Language} & \textbf{Days} &\textbf{Firefox Language}& \textbf{Days}\\\hline
C++& 39.18 & C++& 51.21\\ \hline
JS& 34.28 & JS & 66.47\\ \hline
HTML & 65.87 & HTML& 81.90\\ \hline
C & 31.25 & C & 52.99\\ \hline
XML & 114.00 & XML & 110.81\\ \hline
Python&73.71&Python&87.57\\ \hline
Java&76.11&Java&189.70\\ \hline
CSS&69.44&CSS&330.42\\ \hline
\end{tabular}
}
\end{table}

\cref{table:AvgPatchingOtherFeatures} shows how average patching time varies with severity, weakness type, components, and programming languages.

\begin{table}[!ht]
\centering
\caption{Chi-Squared Test Results}
\label{table:testresults}
\vspace{-1em}
\resizebox{\columnwidth}{!}{
\begin{tabular}{|c|r||c|r|}
\hline
\textbf{Chromium Internal vs. External Reports} &\textbf{$p$-Value} & \textbf{Firefox Internal vs. External Reports} &\textbf{$p$-Value}\\
\hline
Impacted Releases & $<.001$& Impacted Releases& $<.001$ \\ \hline
Security-Severity & $<.001$ &Security-Severity&  $<.001$\\ \hline
Component & 0.0 &Component& $<.001$\\ \hline
Weakness Types &$<.001$ &Weakness Types&$<.001$\\ \hline
Language & $<.001$ &Language&$<.001$\\ \hline
\hline
\textbf{Chromium Internal vs. External Reports (Only Stable)} &\textbf{$p$-Value} & \textbf{Firefox  Internal vs. External Reports (Only Stable)} &\textbf{$p$-Value}\\
\hline
Security-Severity & $<.001$ &Security-Severity& $<.001$ \\ \hline
Component & 0.0 &Component& $<.001$\\ \hline
Weakness Types &$<.001$ &Weakness Types& $<.001$\\ \hline
Language & $<.001$ &Language&$<.001$\\ \hline
\hline
\textbf{Chromium Rediscoveries} &\textbf{$p$-Value} & \textbf{Firefox Rediscoveries} &\textbf{$p$-Value}\\ \hline
Impacted Releases & $<.001$& Impacted Releases& 0.006 \\ \hline
Security-Severity & $<.001$ &Security-Severity& $<.001$\\ \hline
Component & 0.0 &Component& 0.0\\ \hline
Weakness Types &0.0 &Weakness Types&$<.001$\\ \hline
Language & 0.0&Language&0.0\\ \hline \hline
\textbf{Chromium Exploited vs. All Other Vulnerabilities} & \textbf{$p$-Value} &\textbf{Firefox Exploited vs. All Other Vulnerabilities} &\textbf{$p$-Value}\\ \hline
Impacted Releases &$<.001$& Impacted Releases& 0.001\\ \hline
Security-Severity & 0.06 &Security-Severity& 0.006 \\ \hline
Component & 0.04 &Component& $<.001$\\ \hline
Weakness Types &0.87 &Weakness Types&0.99\\ \hline
Language & 0.45&Language &0.97\\ \hline
\hline
\textbf{Chromium Exploited vs. All External Vulnerabilities} & \textbf{$p$-Value} &\textbf{Firefox Exploited vs. All External Vulnerabilities} &\textbf{$p$-Value}\\ \hline
Impacted Releases &0.01& Impacted Releases&0.01  \\ \hline
Security-Severity &  0.01&Security-Severity&0.003 \\ \hline
Component &  &Component& \\ \hline
Weakness Types &&Weakness Types&\\ \hline
Language &0.21&Language &\\ \hline
\end{tabular}
} 
\end{table}

\cref{table:testresults} shows the results of chi-squared tests between different types of vulnerabilities. For some variables, we could not apply tests due to the 0 values (empty cells).    

\section{Internal and External Reports Impacting Stable Releases}
\label{InExStable}

\pgfplotstableread[col sep=comma,header=true]{data/Chrom/ie/Chrom_ie_stable_Broad_type.csv}\ChromBroadsIeStable
\pgfplotstableread[col sep=comma,header=true]{data/Chrom/ie/Chrom_ie_stable_CompGroup.csv}\ChromComponentsIeStable

\pgfplotstableread[col sep=comma,header=true]{data/Chrom/ie/Chrom_ie_stable_Languages.csv}\ChromLangsIeStable

\pgfplotstableread[col sep=comma,header=true]{data/Firefox/ie/Firefox_ie_stable_Broad_type.csv}\FirefoxBroadsIeStable
\pgfplotstableread[col sep=comma,header=true]{data/Firefox/ie/Firefox_ie_stable_Component.csv}\FirefoxComponentIeStable

\pgfplotstableread[col sep=comma,header=true]{data/Firefox/ie/Firefox_ie_stable_Language.csv}\FirefoxLangIeStable

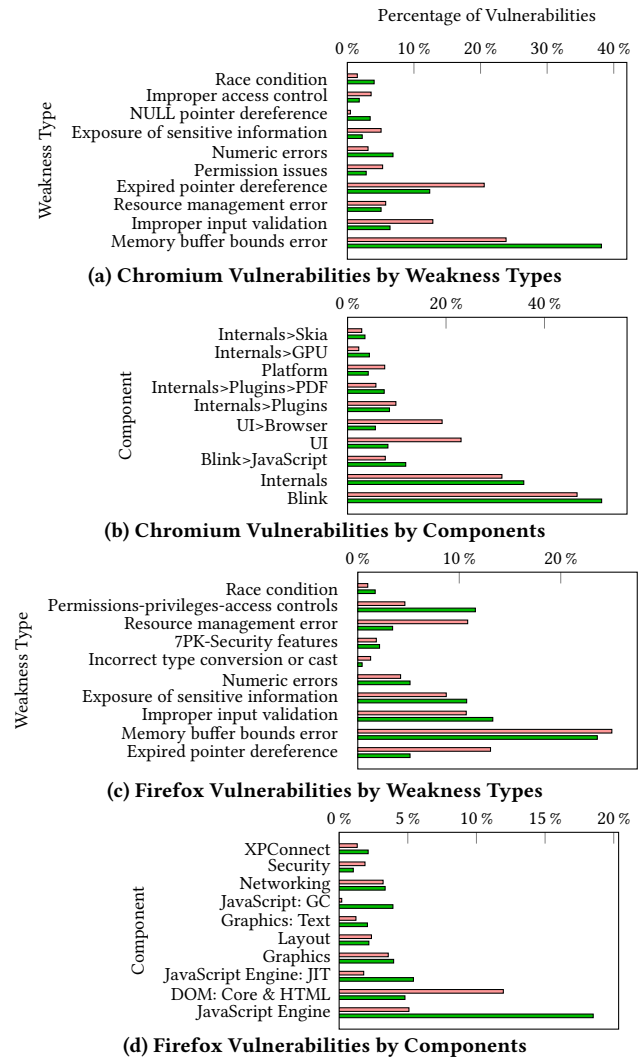
\begin{figure}
\begin{subfigure}[b]{\linewidth}
\null\hfill\begin{tikzpicture}[font=\footnotesize]
\null\vspace{1em}%
\begin{axis}[
    width=0.625\textwidth,
    height = 4.2cm, 
    xbar, 
    xmin=0, %
    ylabel style={align=center},
    ylabel=Weakness Type,
    xlabel style={align=center},
    xlabel=Percentage of Vulnerabilities,
    ytick=data,
     ytick style={draw=none},
    legend pos=north east,
    legend columns=2,
    xticklabel=\pgfmathprintnumber{\tick}\,$\%$,
    xtick pos=upper,xticklabel pos=upper,
    yticklabels from table={\ChromBroadsIeStable}{Broad_type}
    ]
    \addplot [GreenBars,yshift=0.5pt] table[ y expr=\coordindex, x expr=\thisrow{internalPercentage}] {\ChromBroadsIeStable};
    \addplot [RedBars,yshift=-0.5pt] table[ y expr=\coordindex, x expr=\thisrow{externalPercentage}] {\ChromBroadsIeStable};
\end{axis}
\end{tikzpicture}
\null\vspace{-0.5em}%
\caption{Chromium Vulnerabilities by Weakness Types}
\label{subfig:ChromBroadsIeStable}
\end{subfigure}
\null\vspace{-1.5em}
\begin{subfigure}[b]{\linewidth}
\null\hfill\begin{tikzpicture}[font=\footnotesize]
\begin{axis}[
    xbar, xmin=0,width=0.625\textwidth,
    height = 4.2cm, 
    ylabel=Component,
    ytick=data,
     ytick style={draw=none},
    legend pos=north east,
    legend columns=2,
    xticklabel=\pgfmathprintnumber{\tick}\,$\%$,
    xtick pos=upper,xticklabel pos=upper,
    yticklabels from table={\ChromComponentsIeStable}{CompGroup},
    ]
    \addplot [GreenBars,yshift=0.5pt] table[ y expr=\coordindex, x expr=\thisrow{internalPercentage}] {\ChromComponentsIeStable};
    \addplot [RedBars,yshift=-0.5pt] table[ y expr=\coordindex, x expr=\thisrow{externalPercentage}] {\ChromComponentsIeStable};
\end{axis}
\end{tikzpicture}
\null\vspace{-0.5em}
\caption{Chromium Vulnerabilities by Components}
\label{subfig:ChromComponentIeStable}
\end{subfigure}
\null\vspace{-1.5em}%

\begin{subfigure}[b]{\linewidth}
\null\hfill\begin{tikzpicture}[font=\footnotesize]
\begin{axis}[
    width=0.625\textwidth,
    height = 4.2cm, 
    xbar, 
    xmin=0, %
    ylabel style={align=center},
    ylabel=Weakness Type,
    xlabel style={align=center},
    ytick=data,
    ytick style={draw=none},
    legend pos=north east,
    legend columns=2,
    xticklabel=\pgfmathprintnumber{\tick}\,$\%$,
    xtick pos=upper,xticklabel pos=upper,
    yticklabels from table={\FirefoxBroadsIeStable}{Broad_type}
    ]
    \addplot [GreenBars,yshift=0.5pt] table[ y expr=\coordindex, x expr=\thisrow{internalPercentage}] {\FirefoxBroadsIeStable};
    \addplot [RedBars,yshift=-0.5pt] table[ y expr=\coordindex, x expr=\thisrow{externalPercentage}] {\FirefoxBroadsIeStable};
\end{axis}
\end{tikzpicture}
\null\vspace{-1.5em}
\caption{Firefox Vulnerabilities by Weakness Types}
\label{subfig:FirefoxBroadtypessIeStable}
\end{subfigure}
\null\vspace{-1.5em}
\begin{subfigure}[b]{\linewidth}
\null\hfill\begin{tikzpicture}[font=\footnotesize]
\begin{axis}[
    legend style={at={(0.485, 0.99)},anchor=north east},
    xbar, xmin=0,width=0.625\textwidth,
    height = 4.2cm, 
    ylabel=Component,
    ytick=data,
    ytick align=inside,
    ytick style={draw=none},
    legend pos=north east,
    legend columns=2,
    xticklabel=\pgfmathprintnumber{\tick}\,$\%$,
    xtick pos=upper,xticklabel pos=upper,
    yticklabels from table={\FirefoxComponentIeStable}{Component},
    ]
    \addplot [GreenBars,yshift=0.5pt] table[ y expr=\coordindex, x expr=\thisrow{internalPercentage}] {\FirefoxComponentIeStable};
    \addplot [RedBars,yshift=-0.5pt] table[ y expr=\coordindex, x expr=\thisrow{externalPercentage}] {\FirefoxComponentIeStable};
\end{axis}
\end{tikzpicture}
\null\vspace{-0.5em}
\caption{Firefox Vulnerabilities by Components}
\label{subfig:FirefoxComponentIeStable}
\end{subfigure}
\null\vspace{-1.5em}

\caption{Comparison of internal (\coloredsquare{ColorLegendGreen}) and external (\coloredsquare{ColorLegendRed}) security reports in \emph{stable} releases of Chromium and Firefox.}
 \label{fig:ieStable}
\end{figure}

\pgfplotstableread[col sep=comma]{data/Chrom/exploited/Chrom_ie_exploited_otherBroad_type.csv}\ChromOtherBroad
\pgfplotstableread[col sep=comma]{data/Chrom/exploited/Chrom_ie_exploited_otherLanguages.csv}\ChromOtherLang
\pgfplotstableread[col sep=comma]{data/Chrom/exploited/Chrom_ie_exploited_otherComponent.csv}\ChromOtherComponent

\pgfplotstableread[col sep=comma]{data/Firefox/exploited/Firefox_exploited_other_Component.csv}\FirefoxOtherComponent
\pgfplotstableread[col sep=comma]{data/Firefox/exploited/Firefox_exploited_other_Broad_type.csv}\FirefoxOtherBroad
\pgfplotstableread[col sep=comma]{data/Firefox/exploited/Firefox_exploited_other_Language.csv}\FirefoxOtherLanguages

\begin{figure}

\begin{subfigure}[b]{\linewidth}
\null\hfill\begin{tikzpicture}[font=\footnotesize]
\begin{axis}[
    width=0.625\textwidth,
    height = 3.2cm, 
    xbar, 
    xmin=0, %
    ylabel style={align=center},
    ylabel=Weakness Type,
    xlabel style={align=center},
    xlabel=Percentage of Vulnerabilities,
    ytick=data,
       ytick align=inside,
    ytick style={draw=none},
    legend pos=north east,
    legend columns=2,
    xticklabel=\pgfmathprintnumber{\tick}\,$\%$,
    xtick pos=upper,xticklabel pos=upper,
    yticklabels from table={\ChromOtherBroad}{Broad_type}
    ]
    \addplot [PurpleBars,yshift=0.5pt] table[ y expr=\coordindex, x expr=\thisrow{exploitedPercentage}] {\ChromOtherBroad};
    \addplot [BrownBars,yshift=-0.5pt] table[ y expr=\coordindex, x expr=\thisrow{otherPercentage}] {\ChromOtherBroad};
\end{axis}
\end{tikzpicture}
\null\vspace{-0.5em}
\caption{Chromium Vulnerabilities by Weakness Types}
\label{subfig:ChromBroadsOther}
\end{subfigure}
\null\vspace{-1.5em}
\begin{subfigure}[b]{\linewidth}
\null\hfill\begin{tikzpicture}[font=\footnotesize]
\begin{axis}[
    legend style={at={(0.485, 0.99)},anchor=north east},
    xbar, xmin=0,width=0.625\textwidth,
    height = 4.2cm, 
    ylabel=Component,
    ytick=data,
    xtick={0, 20, 40},
    ytick align=inside,
    ytick style={draw=none},
    legend pos=north east,
    legend columns=2,
    xticklabel=\pgfmathprintnumber{\tick}\,$\%$,
    xtick pos=upper,xticklabel pos=upper,
    yticklabels from table={\ChromOtherComponent}{CompGroup},
    ]
    \addplot [PurpleBars,yshift=0.5pt] table[ y expr=\coordindex, x expr=\thisrow{exploitedPercentage}] {\ChromOtherComponent};
    \addplot [BrownBars,yshift=-0.5pt] table[ y expr=\coordindex, x expr=\thisrow{otherPercentage}] {\ChromOtherComponent};
\end{axis}
\end{tikzpicture}
\null\vspace{-0.5em}
\caption{Chromium Vulnerabilities by Components}
\label{subfig:ChromComponentOther}
\end{subfigure}
\null\vspace{-1.5em}%

\begin{subfigure}[b]{\linewidth}
\null\hfill\begin{tikzpicture}[font=\footnotesize]

\begin{axis}[
    width=0.625\textwidth,
    height = 3.2cm, 
    xbar, 
    xmin=0, %
    ylabel style={align=center},
    ylabel=Weakness Type,
    xlabel style={align=center},
    ytick=data,
     ytick align=inside,
    ytick style={draw=none},
    legend pos=north east,
    legend columns=2,
    xticklabel=\pgfmathprintnumber{\tick}\,$\%$,
    xtick pos=upper,xticklabel pos=upper,
    yticklabels from table={\FirefoxOtherBroad}{Broad_type}
    ]
    \addplot [PurpleBars,yshift=0.5pt] table[ y expr=\coordindex, x expr=\thisrow{exploitedPercentage}] {\FirefoxOtherBroad};
    \addplot [BrownBars,yshift=-0.5pt] table[ y expr=\coordindex, x expr=\thisrow{otherPercentage}] {\FirefoxOtherBroad};
\end{axis}
\end{tikzpicture}
\null\vspace{-0.5em}
\caption{Firefox Vulnerabilities by Weakness Types}
\label{subfig:FirefoxBroadsOther}
\end{subfigure}
\null\vspace{-3em}%
\begin{subfigure}[b]{\linewidth}
\null\hfill\begin{tikzpicture}[font=\footnotesize]
\begin{axis}[
    xbar, xmin=0,width=0.625\textwidth,
    height = 4.2cm, 
    ylabel=Component,
    ytick=data,
    xtick={0, 20, 40},
      ytick align=inside,
    ytick style={draw=none},
    legend pos=north east,
    legend columns=2,
    xticklabel=\pgfmathprintnumber{\tick}\,$\%$,
    xtick pos=upper,xticklabel pos=upper,
    yticklabels from table={\FirefoxOtherComponent}{Component},
    ]
    \addplot [PurpleBars,yshift=0.5pt] table[ y expr=\coordindex, x expr=\thisrow{exploitedPercentage}] {\FirefoxOtherComponent};
    \addplot [BrownBars,yshift=-0.5pt] table[ y expr=\coordindex, x expr=\thisrow{otherPercentage}] {\FirefoxOtherComponent};
\end{axis}
\end{tikzpicture}
\null\vspace{-0.5em}
\caption{Firefox Vulnerabilities by Components}
\label{subfig:FirefoxComponentOther}
\end{subfigure}
\null\vspace{1em}%

\caption{Comparison of exploited vulnerabilities (\coloredsquare{ColorLegendPurple}) and all other vulnerabilities (\coloredsquare{ColorLegendBrown}) in Chromium and Firefox.}
\label{fig:ExploitedIssues_vs_AllOtherIssues}
\end{figure}
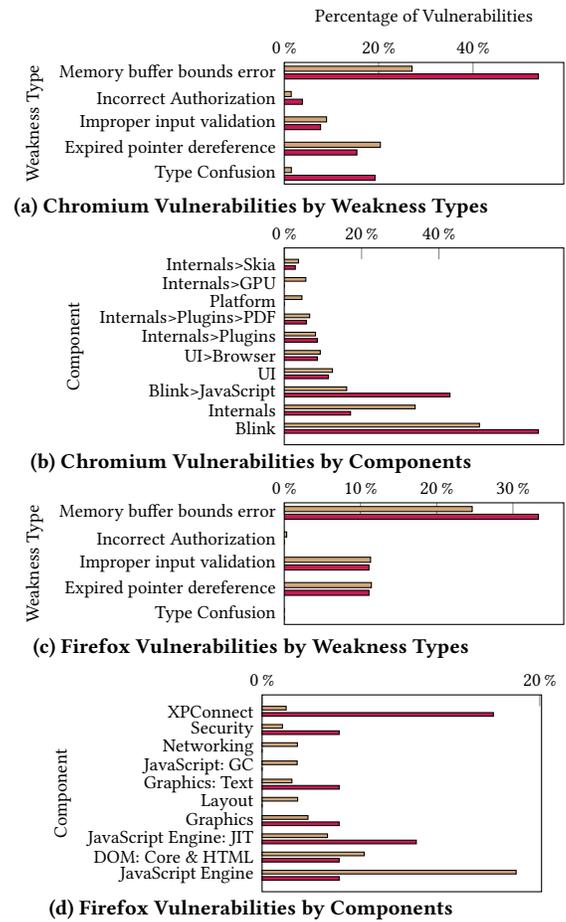

\cref{fig:ieStable} shows the distribution of weakness types and impacted components for internal and external reports in stable releases. As for weakness types, we find that reports related to \emph{Memory buffer bounds error} are the most common among both origins, in both Chromium and Firefox (\cref{subfig:ChromBroadsIeStable,subfig:FirefoxBroadtypessIeStable}). In Chromium, the \emph{Blink} component is most commonly impacted by both internal and external reports (\cref{subfig:ChromComponentIeStable}). In Firefox, \emph{Javascript Engine} is most common among internal reports, while \emph{Dom: Core \& HTML} is most common among external ones (\cref{subfig:FirefoxComponentIeStable}). 
We also compared internal and external reports in terms of severity and programming languages. Most internal and external reports have \emph{high severity} in both Chromium and Firefox. External reports are more common than internal ones among vulnerabilities with \emph{critical severity} in both software products. 
As for programming languages, we find that most reports are related to  \emph{C++}, regardless of origins, for both Chromium and Firefox. 
Pearson’s chi-squared test shows that external and internal reports (that impact stable releases)  follow significantly different distributions in terms of severity, weakness type, impacted components, and programming languages in both Chromium and Firefox.

\section{Comparison of Exploited Vulnerabilities}
\label{ExploitedOtherAppendix}

We compare vulnerabilities that are exploited in the wild with all other vulnerabilities (i.e., vulnerabilities that have not been exploited).
\cref{fig:ExploitedIssues_vs_AllOtherIssues} shows the distributions of weakness types and components for exploited vulnerabilities and all other vulnerabilities for both Chromium and Firefox. 
\ifExtendedVersion
Regarding release channel, we observe that 100\% of exploited Chromium vulnerabilities and 88.88\% of exploited Firefox vulnerabilities impacted \emph{stable} release channel whereas 49.37\% and 60.40\% of all other vulnerabilities that have not been exploited pertain to \emph{stable} releases in Firefox and Chromium respectively. As for security-severity 71.42\% of exploited Chromium vulnerabilities were labeled as \emph{high} severity whereas 49.92\% of all other vulnerabilities have \emph{high} severity. In Firefox 62.50\% of exploited vulnerabilities have \emph{critical} severity while 25.65\% of all other vulnerabilities have \emph{critical} severity assigned to them.
Among programming languages, \emph{C++} has the highest exploited reports among other languages in both datasets. This also holds for all other vulnerabilities that have not been exploited. 

Comparison of exploited vulnerabilities and other vulnerabilities in terms of weakness type shows that vulnerabilities with \emph{Memory buffer bound error} have a higher percentage in both exploited vulnerabilities and other vulnerabilities in both Chromium and Firefox datasets (see \cref{subfig:ChromBroadsOther} and \cref{subfig:FirefoxBroadsOther}). For component attribute in Chromium (see \cref{subfig:ChromComponentOther}), vulnerabilities with \emph{Blink} component have a higher percentage in both exploited vulnerabilities and other vulnerabilities. In Firefox, vulnerabilities with \emph{XPConnect} component have been exploited more than vulnerabilities with other components. For all other vulnerabilities, vulnerabilities with \emph{JavaScript Engine} component has a higher percentage in comparison with others (see \cref{subfig:FirefoxComponentOther}).
\fi
The results of Pearson’s chi-squared test show that exploited vulnerabilities and all other reported vulnerabilities follow significantly different distributions in terms of impacted release channels and components in Chromium. Chi-squared tests for Firefox show that exploited vulnerabilities and all other reported vulnerabilities follow significantly different distributions in terms of impacted release channels, security severity, and impacted components (weakness types and languages accepted the~null).

\ifExtendedVersion
\section{Data Collection}
\label{sec:data_collection}

In this section, we describe the key steps of our data collection process for both Chromium and Firefox. 

\subsection{Chromium Issue Tracker}
\label{sec:collection_tracker}

We collected all data from September 2, 2008 to September 13, 2022 from the Chromium issue tracker\footnote{\url{https://bugs.chromium.org/p/chromium/issues/}} using Monorail API version 2\footnote{\url{https://chromium.googlesource.com/infra/infra/+/master/appengine/monorail/api/README.md}}. We use three types of requests from the Monorail API for data collection: \begin{enumerate*}[label=(\roman*)]
\item \textit{ListIssues}, which returns the list of reports that satisfy the query specified in the request;
\item \textit{GetIssue}, which returns the details of the report corresponding to the report identification number specified in the request; and
\item \textit{ListComments}, which returns the list of comments posted on the report specified in the request.
\end{enumerate*}
For each vulnerability's report, the Chromium issue tracker stores a list of comments, which includes conversations among internal employees and external parties as well as a history of changes (i.e., amendments) made to the report.

Each report contains the following fields:
\begin{itemize}
    \item \emph{IdentificationNumber}: A unique number that identifies the report.
    
    \item \emph{Status}: Current state of the report (\emph{Unconfirmed}, \emph{Untriaged}, \emph{Invalid}, \emph{Duplicate}, \emph{Available}, \emph{Assigned}, \emph{Started}, 
    \emph{ExternalDependency},
    \emph{Fixed}, \emph{Verified}, and \emph{Archived}).
    
    \item \emph{Component}: Component (or components) of the Chromium project that are affected by the report.

    \item \emph{Owner}: Email address of the person who currently owns the report (e.g.,  reporter of the vulnerability or the person who fixes or closes the vulnerability).
    
    \item \emph{AllLabels}: Labels associated with the report. These labels are used to categorize reports, e.g., to indicate
    security-severity levels (\emph{critical}, \emph{high}, \emph{medium}, or \emph{low}), 
    impacted versions (\emph{stable}, \emph{beta}, or \emph{head}), 
    reward amount for bug bounty (e.g., \snippet{Reward-500} indicates that \$500 is awarded to the reporter of the vulnerability),
    or CVE ID.
    
    \item \emph{Summary}: Short description of the report.
    
    \item \emph{ReporterEmail}: Email address of the person who reported the report. 
    
    \item \emph{CCDetails}: Email addresses of all users who are part of the conversation thread of the report.

    \item \emph{OpenedTimestamp}: Date and time when the report was initially reported.
    
    \item \emph{ClosedTimestamp}: Date and time when the report was closed.
    
    \item \emph{BugType}: Type of the report (e.g., \snippet{Bug-Security}).
    
    \item \emph{MergedInto}: This is an optional field that applies only to duplicate reports. This field references the original report.
\end{itemize}

\noindent
Each comment posted on a report consists of the following fields:
\begin{itemize}
    \item \emph{CommenterEmail}: Email address of the person who posted the comment.
    
    \item \emph{Content}: Text of the comment, images of the report, videos of how to reproduce the report, etc.
    
    \item \emph{SequenceNumber}: Order of the comment among all comments posted on the report.
    
    \item \emph{CommentedTimestamp}: Date and time when the comment was posted on the report.
        
    \item \emph{Amendments}: Updating or removing the values of some fields of the report
    (e.g., changing the owner, status, or priority).
\end{itemize}

\noindent
Each amendment added to a comment consists of the following fields:
\begin{itemize}
    \item \emph{FieldName}: Name of the field that the amendment changes.
   
    \item \emph{OldValue}: List of previous values of the field. This an optional field.
    
    \item \emph{NewOrDeltaValue}: List of new  and removed values of the field.  
    
    \item \emph{AmendmentTimestamp}: Date and time when the amendment was posted.
\end{itemize}

\paragraph{Chrome Releases}
\label{subsubsec:chrome_releases}
We also collected all the release notes from the Chrome Releases blog\footnote{\url{https://chromereleases.googleblog.com/}}, which provides information regarding both the closed-source Chrome and the open-source Chromium projects.  A release note is a  blog post written by Google when they officially release a new version of Chrome or Chromium.
Each Chromium release note contains a list of vulnerabilities that Google patches in the Chromium project when releasing the corresponding version. Each entry in this list of vulnerabilities contains the following fields:
\begin{itemize}
    \item \emph{IdentificationNumber}: Unique identification number of a report. We use this to join the Chromium issue tracker data with the Chrome Releases dataset.
    \item \emph{ReporterName}: List of bug hunters who reported the particular vulnerability.
    \item \emph{Association}: Organization (or organizations) where the reporters work.
    \item \emph{ReleaseDate}: Release date of Chromium version that includes the fix for the vulnerability. 
 \end{itemize}

\paragraph{Google Git For Programming Languages Analysis}
Another data resource that we use in our study is the Google Git repository\footnote{\url{https://chromium.googlesource.com/}}. From analyzing comments on vulnerabilities that we collected from the Chromium issue tracker, we find that most vulnerabilities that have been fixed have links to the Google Git repository, which we can use to identify the files that were changed to fix the vulnerability. For each vulnerability with a Google Git repository link, we collected the programming languages of the files that were changed.

\subsection{Mozilla Firefox VRP}
We collected data from two main sources, Bugzilla \footnote{\url{https://bugzilla.mozilla.org/home}} (Firefox bug tracker) and Known Vulnerabilities from the Mozilla website \footnote{\url{https://www.mozilla.org/en-US/security/known-vulnerabilities/}}, which is a list of security advisories based on product and advisories for older products that all are listed in the Mozilla Foundation Security Advisories (MFSA) \footnote{\url{https://www.mozilla.org/en-US/security/advisories/}}. We collected all security issues from January 24, 2012 to August 25, 2022. 

To collect security vulnerabilities, we used security keywords added to the URL of Bugzilla search and scraped all URLs of reports that had at least one of the security keywords in the report \emph{Keywords} field. Finally, all of the information related to a report was scraped. 
Each report contains several fields, of which the collected ones are listed below:
\begin{itemize}
    \item \emph{BugID}: A unique identifier of the report.
     \item \emph{CVE}: CVE Id of the report (does not exist for all of the reports).
    \item \emph{Opened}: Date and time when the report is opened.
    \item \emph{Closed}: Date and time when the report is closed.
    \item \emph{Summary}: A brief summary of  the report.
    \item \emph{Product}: Product type which the report is related to (we are interested in the Core and Firefox).
     \item \emph{Component}: Component (or components) of the Firefox that are affected by the vulnerability. 
      \item \emph{Type}: This field represents type of the bug. It contains three types: defect, enhancement, and task. We are only interested in the defect type.
     \item \emph{Status}: This represents current status of the report and what has happened to the report. It contains UNCONFIRMED, NEW, ASSIGNED, REOPENED for reports that are open. For reports that are closed, \emph{Status} field contain, RESOLVED and VERIFIED which each have 7 resolutions (resolution represents the approach applied to reach to the current status): FIXED, INVALID, WONTFIX, MOVED, DUPLICATE, WORKSFORME, and INCOMPLETE.
     \item \emph{Reporter} Username of a person who reported the issue.
     \item \emph{Keywords} Criticality of the report (critical, high, medium, low) is mentioned in this field.
      \item \emph{Duplicates} ID number of duplicates of the report.
    \item \emph{Whiteboard} It contains tags, or information of a report's status.  \emph{reporter-external} tag is one of the tags which is used to identify external versus internal reporter. 
     \item \emph{Bug Flags} It contains \emph{sec-bounty} value and it does not exist for all of the reports.
      \item \emph{TrackingFlagsStatus} It contains vulnerability's statuses tracked by developers (does not exist for all of the reports).
      \item \emph{Comments} All comments posted on the report.
\end{itemize}  
\paragraph{Fixed Timestamp}
Each report that has \emph{VERIFIED} or \emph{RESOLVED} followed by \emph{FIXED} in its \emph{Status} field, has an arrow followed by \emph{RESOLVED} ($\rightarrow$ RESOLVED) and an arrow followed by \emph{FIXED} ($\rightarrow$ FIXED) in one of its last comments (date of that comment which equals to the close time of the report). We use that date (close time) as the date the vulnerability is fixed. There are some reports that are closed because of incomplete fix status and reopened again. In those cases, we consider the last fixed time as the fix time of that vulnerability.

\paragraph{Mozilla Foundation Security Advisories (MFSA)}
MFSA reports vulnerabilities for Mozilla products. In this paper, we focus on Firefox.
To identify reports pertaining to \emph{stable} releases, we use MFSA. For FireFox and its older versions, we scraped advisories to be able to label reports that pertain to stable releases.
We also collected the \emph{Reporter} field, which some pages in MFSA have, to identify external versus internal reporters in our cleaning process.

\paragraph{Firefox Modified Source Files For Programming Languages Analysis}
Most vulnerabilities that have been fixed have links to the Mozilla source-code repositories in their comments, which we use to identify the files that were changed to fix the vulnerability. For each vulnerability with a %
repository link, we collect the programming languages of the files that were changed. 

\paragraph{Reopened}
In Firefox, some reports had incomplete status due to the lack of information for replication and patching. For some of them, Mozilla reopened a new report of the vulnerability, which was then completed with respect to this information, and marked the first report as a duplicate. These reports can be identified by searching a right arrow to \emph{REOPENED} ($\rightarrow$ REOPENED) in the comments of that reports.
Later in our analysis, we exclude these reports from rediscovery analysis since they are not actual rediscoveries.

\subsection{CVEs and CWEs}
\paragraph{CVEDetails}
We leverage CVEDetails\footnote{\url{https://www.cvedetails.com/}} and MITRE CWE\footnote{\url{https://cwe.mitre.org/}} to collect information regarding CVE IDs and weakness types (CWEs), for both Chromium and Firefox when available.
One of the fields associated with a report is \emph{AllLabels} in Chromium. These labels may include a categorical parameter \emph{Common Vulnerabilities and Exposures (CVE) Entry}, which contains an identification number called \emph{CVE ID}. In Firefox, some reports contain \emph{CVE ID} field which we collected them for analysis.
These identifiers are used by cybersecurity product and service vendors and researchers as one of the standard methods for identifying publicly known vulnerabilities and for cross-linking with other repositories that also use CVE IDs.
Using these CVE IDs, we collected CVSS scores, impact metrics, and weakness types from CVEDetails\footnote{\url{https://www.cvedetails.com/}}. For each report with a CVE ID, we collected the following details: %
\begin{itemize}
    \item \emph{CVSS Score}: \emph{Common Vulnerability Scoring System} (CVSS) provides a way of capturing the fundamental characteristics of a vulnerability. This numerical score reflects the severity of the vulnerability.
    
    \item \emph{Confidentiality Impact}: Impact of successful exploitation on information access and disclosure. %
    
    \item \emph{Integrity Impact}: Impact of successful exploitation on the trustworthiness and veracity of information. %
    
    \item \emph{Availability Impact}: Impact of successful exploitation on the accessibility of information resources.  %
    
    \item \emph{Access Complexity}: Complexity of the attack required to exploit the vulnerability once an attacker has gained access to the system.
    
    \item \emph{CWE ID}: \emph{Common Weakness Enumeration} (CWE) is a community-developed list of weakness types.
    The CWE ID references the type of software weakness associated with the particular vulnerability.
\end{itemize}

\paragraph{Weakness Types (CWE IDs)}
\label{sub_sub_sec:CWEID}
The CWE IDs associated with the vulnerabilities represent common types of software weaknesses. We later use CWE ID collected from \emph{cvedetails} to collect broad-type names of CWEs from MITRE for weakness type analyses. 
Some of these weakness types have a hierarchical relationship with other types. For example, CWE 119 denotes the error ``Improper Restriction of Operations within the Bounds of a Memory Buffer.'' This weakness type is also the parent of other CWEs, including CWE 120 (Classic Buffer Overflow), CWE 125 (Out-of-bounds Read), and CWE 787 (Out-of-bounds Write). For ease of presentation, we group the CWE weakness types together based on their parent-children hierarchy.

\section{Data Cleaning}
\label{sec:data_cleaning}

In this section, we describe the key steps of our data cleaning process for both Chromium and Firefox. 

In our analysis, we only consider reports that satisfy at least one of the following three conditions: (1) the report is an original report, and it has at least one security label; (2) the report is an original report, and the value of field BugType is \emph{Bug-Security} for Chromium or has one of security labels in \emph{Keywords} field for Firefox; and (3) the report is a duplicate report,
and its original report satisfies at least one of the above conditions. 
\subsection{Duplicate Reports}

\label{sub_sec:duplicate}

\subsubsection{Chromium} 

A report in the issue tracker is considered to be a \emph{duplicate} if the underlying report has already been reported to the Chromium issue tracker (i.e., if this is a rediscovery). We can determine whether a report is a duplicate or not based on the \emph{Status} field of the report: if the \emph{Status} field is marked as \emph{Duplicate}, the report is a duplicate. 

To facilitate studying vulnerability rediscovery, we find the original report of each duplicate report as follows.
For each duplicate report~$D$, we follow the \emph{MergeInto} field to retrieve the report referenced by it. If that is a duplicate report, we again follow the \emph{MergeInto} field of the referenced report. We continue this process recursively until either one of the following holds:
\begin{itemize}
    \item We reach a report $O$ that is not a duplicate report. In this case, report $O$ is the \emph{original report} of duplicate report~$D$. 
    \item We reach a report $X$ that is a duplicate report but does not have any references in the \emph{MergedInto} field (or the value of \emph{MergedInto} field is malformed). In this case, we say that the duplicate report $D$ does not have an original report. 
\end{itemize}

We include a duplicate report $D$ in our rediscovery analysis if report $D$ has an {original report}~$O$ and report $O$ has at least one security-related label. In order to retrieve duplicates of a vulnerability, we use \emph{Duplicates} field which has references to the duplicate reports. For the cases that references also have a reference to other duplicates, we recursively retrieve duplicate reports until there is not any reference to a duplicate report.
\subsubsection{Firefox}

We can determine whether a vulnerability was reported earlier (i.e., is a duplicate) or not based on the \emph{Status} field. If the report is a duplicate, \emph{Status} field contains the keyword \emph{Duplicate} and it has reference to the original report. In some cases, the referenced report, which is supposed to be the original report, has \emph{Duplicate} in the status and has reference to another report. In these cases, we recursively, retrieve the report which is referenced in the \emph{Status} field until there is not any reference to a report. 

\subsection{Valid and Invalid Reports}
\label{sub_sec:valid_invalid}
For both Chromium and Firefox, if the \emph{Status} field of an original report is not marked as \emph{Invalid}, it is considered a \emph{valid original report}. If a duplicate report has a valid original report, then the duplicate report is a \emph{valid duplicate report}. If a vulnerability belongs to either valid original reports or valid duplicate reports, then the vulnerability is considered a \emph{valid vulnerability}.

If the \emph{Status} field of an original report is marked as \emph{Invalid}, it is considered an \emph{invalid original report}. If a duplicate report has an invalid original report, then the duplicate report is an \emph{invalid duplicate report}. If a report belongs either to invalid original reports or invalid duplicate reports, then, the report is considered an \emph{invalid report}.

In Firefox, there are other invalid statuses that we do not consider them as valid statuses for reports. We do not consider duplicate reports that their original report has \emph{INACTIVE}, \emph{INCOMPLETE}, \emph{MOVED}, \emph{WONTFIX}, \emph{WORKSFORME}, or \emph{UNCONFIRMED} in its' \emph{Status} field. However, there is an exception here. By checking some of the reports with the mentioned statuses, we realized that some reports that have  \emph{WORKSFORME} or \emph{INCOMPLETE} in their \emph{Status} field, have \emph{fixed} word in their \emph{TrackingFlags} field. Therefore, we keep duplicate reports that their original has \emph{WORKSFORME} or \emph{INCOMPLETE} in its' status and it got fixed in a version (according to the 'fixed' word in the \emph{TrackingFlags} field).

\subsection{Type and Product}

In Bugzilla (Firefox), there is a \emph{Type} field that contains type of a vulnerability which can be task, enhancement, or defect. We only keep duplicate reports that their original report have \emph{defect} type. 
As for \emph{Product} field, we keep only duplicate reports that their original report's product contains \emph{Core} or \emph{Firefox}.
\subsection{External and Internal Reports}
\subsubsection{Chromium}

\label{sub_sec:extern_vs_internal}
The Chromium issue tracker contains reports either reported internally by Google or reported externally by bug hunters. For each report, we use the reporter's email address to classify the report as either an \emph{internal} or an \emph{external report}. However, not all email addresses fall into the internal vs. external classification; thus, we cannot always determine the reported origin based on the email address alone. For each such address, we manually check the activities of the email address, such as vulnerabilities reported and comments posted by this particular email address. We determine the reporter's origin based on the activities associated with a particular email address.

There are also cases where the email address could be misleading. First, some external bug hunters submit reports privately to Google, and internal experts then post these reports on the Chromium issue tracker. Second, sometimes internal reporters import reports from other bug-bounty programs (e.g., Firefox, Facebook). In these cases, we need to identify the actual external reporter for each replicated report by analyzing the CC email address list of the report.  We further improve the data cleaning process of distinguishing internal and external reports using the data we collected from Chrome Releases. The detailed cleaning process can be described using the following steps.

\vspace{0.5em}
\noindent \textbf{Step 1: Initial Classification based on ReporterEmail Field}

For each report \textit{I}, we use the email address of the reporter to classify the report \textit{I} as either an \emph{internal} or an \emph{external} report. Specifically, if the email address is \emph{ClusterFuzz} or ends with \emph{google.com}, \emph{chromium.org}, or \emph{gserviceaccount.com}, and does not contain any label stating \emph{external\_security\_report} or label starting \emph{reward-to-external} and contains a non google email (i.e., email address without google.com or chromium.org) then we consider report \textit{I} to be internally reported; otherwise, we consider it to be externally reported.

\vspace{0.5em}
\noindent \textbf{Step 2: Identifying Outlier Email Addresses based on Comments}

We found that some of the reporters have email addresses that do not fit the rules of Step 1.  One exception is the gmail address \emph{scarybeast}, which belongs to an internal reporter.
We identified this exception by analyzing the comments posted on reports reported by this email address. Based on the comments, we determined that this reporter is one of the key persons in announcing the confirmation of the reward to external reporters. Thus, we consider reports reported by this email address as internal reports. \footnote{We believe this person was actually a member of the Google Chrome Security Team.}

We also find some other exceptions where the email address of the reporter \emph{skylined} or \emph{cnardi} end with chromium.org. When we analyze the comments on reports reported by \emph{skylined} with chromium.org address, we realized he served as a team member of the Google Chrome Security Team from 2008 - 2013 and left Google. After leaving Google, he reported few vulnerabilities as an external reporter and received rewards. When we analyze the comments on reports reported by \emph{cnardi} ends with chromium.org, one comment mentions ``\snippet{cnardi@chromium.org} as an external reporter regardless of his email address ends with @chromium.org.'' Accordingly, we classify them as an external reporter. We consider the reports reported by these two reporters as external reports.

\vspace{0.5em}
\noindent \textbf{Step 3: Analyzing CCDetails Field and Identifying the Actual External Reporters}

Some reports that are submitted by internal-reporters are replications of reports privately submitted by external reporters to Google or reports imported from other bug bounty programs (e.g., Firefox, Facebook). 
Google replicates most of these externally reported reports through the automated tool \texttt{ClusterFuzz}, but sometimes Google replicates them manually using internal reporters (e.g., \snippet{scarybeasts@gmail.com}). 
For each replicated report, we need to identify the actual external reporter. We use the following approach and identify the email address of the external reporter of those reports.

For each report \textit{I} for which we have to identify the email address of the actual external reporter, we first extract the CC email addresses (\textit{CC$_{all}$}) from the \emph{CCDetails} field. From \textit{CC$_{all}$}, we obtain a new list \textit{CC$_{remain}$} by removing the email address where the email address belongs to an internal reporter at the end of Step 2. For each email address in the \textit{CC$_{remain}$} list, we look into comments of the corresponding report \textit{I} whether any comment has one of the following phrases ``originally reported by'', ``thanks to'', ``credits to'', ``thanks'', ``credits'', ``reward'', ``congratulations'' immediately followed by username or email address or full name of the reporter. If the username of the email address or the reporter’s full name matches the email address, we add the particular email address to the possible-reporters list.

We repeat the same process for every email address on the list. After the process finishes, we check the possible-reporters list of the report \textit{I}. If the possible-reporters list is empty, we set the \emph{ReporterEmail} field of the report as empty (there are 50 reports for which we cannot identify the email address of the actual external reporter during this data cleaning process). If the possible-reporters list is not empty, then we set the \emph{ReporterEmail} field of the report with the list of email addresses in the potential-reporters list.

Even though there should be only one reporter for each report (i.e., the length of the potential-reporters list should be one), we observe some reports where multiple reporters are rewarded. This may happen when multiple external bug hunters report the same report to Google (not through the issue tracker). Google replicates these reports by posting a single report on the tracker via an internal reporter. In such cases, we let the reporter's email of the report be a list instead of a single email address. Note that for some of these reports with multiple reporters, we perform an additional verification in Step 4.

\vspace{0.5em}
\noindent \textbf{Step 4: Cleaning based on Chrome Releases Data}

Further, we improve the data cleaning process of internal and external reports based on data collected from Chrome Releases (\cref{subsubsec:chrome_releases}). During the last step (Step 3), we mark the \emph{ReporterEmail} field as empty for the reports where we are unable to determine the actual external reporter.

For each report \textit{I} which marked the \emph{ReporterEmail} field as empty in the last step (Step 3), we look for a data entry \textit{DE} with the \emph{IdentificationNumber} field same as the Identification Number field of report \textit{I}. If a data entry exists in the Chrome Releases dataset, then we set the Report Email of report \textit{I} with the Reporter Name in data entry \textit{DE}. Accordingly, we are able to identify the actual external reporter details of 14 reports.

Further, during the last step (Step 3), we have more than one email address set to the \emph{ReporterEmail} field for 13 reports. For each report \textit{I} in those 13 reports, we look for a data entry \textit{DE} with Identification Number field the same as the \emph{IdentificationNumber} field of report \textit{I}.  If there exists a data entry (\textit{DE}) in the Chrome Releases dataset, then we check the value in the ReporterName field of \textit{DE}; if it indicates a single person, then we update the \emph{ReportEmail} field with the single person. We are able to update 8 out of 13 reports with multiple reporters to the actual external reporters. At the end of this step, we were left with 22 reports to go through an additional cleaning process to identify the original \emph{ReporterEmail} field of the report.

For each reporter name \textit{RN} used as the \emph{ReporterEmail} field of the above 22 reports, we list out the reports ($L_{RN}$) reported by the reporter \textit{RN} based on the Chrome Releases dataset. For each report in $L_{RN}$, we look for report \textit{I}, which has the same Identification Number and Reporter Email in the email address format. If we obtain the report \textit{I}, then we map the reporter name {RN} with the \emph{ReporterEmail} field of report \textit{I}; otherwise, we continue the same process with the next report in the list.

Finally, we repeat Step 1 with the cleaned dataset based on Steps 3 and 4. We identify the email address of the actual external reporter for 98\% of valid external reports.

\subsubsection{Firefox}
Reports in Firefox VRP are reported either internally by Firefox internal team members or by external reporters. 

We use four steps to separate internal reports from external reports. First, we use \emph{Whiteboard} and \emph{bug-flag} fields on the report information page. If a report has \emph{reporter-external} in \emph{Whiteboard} field or \emph{sec-bounty} in \emph{bug-flag} field, we consider that report as an externally reported report; otherwise, it is considered as an internally reported report. However, there are reports that do not have the above keywords in the mentioned fields, they are reported by external reporters. To separate them, we added three more cleaning steps. In the second step, we leverage the snow-balling technique (on the comments, such as ``security@mozilla.org received the following report from'') to identify reports (around 650 reports) reported by internal security members of Mozilla and do not have any bounty tag, but their original reporters are external. In the third step, we check reporters with both internal and external reports  (around 50). We manually check whether they are internal or external (by reading comments and checking their social networking websites). In the last step of the cleaning process, we leverage \emph{Reporter} field in MFSA and match the name of reporters in their profile names (we got each reporter's profile name from Bugzilla) with the name of the reporter mentioned in MFSA. By applying the above steps, we are able to separate internal versus external reports with 97\% accuracy. 

\subsection{Vulnerability Attributes}
\label{generalapproach}

For each duplicate report $D$, we clean security-severity, impacted releases, weakness types, components, and programming language attributes to make them consistent with its original report $O$.  Accordingly, we perform the following cleaning steps:

\begin{itemize}
\item If the duplicate report~$D$ does not list any value for an attribute $A$ and the original report~$O$ of duplicate report~$D$ has a value, then we set the value for attribute $A$ of duplicate report~$D$ to the value of the original report~$O$ 

\item If all duplicate reports of original report~$O$ list the same value and original report~$O$ does not list any value for an attribute $A$, then we set the value for attribute $A$ of original report~$O$ to the value of the duplicate reports.

\end{itemize}

Further explanations for different attributes based on the datasets are in the following sections.
\subsubsection{Security-Severity}
\label{sub_sec:severity_levels}

\paragraph{Chromium}
There are four severity levels in the Chromium issue tracker, which describe the security severity of a vulnerability: \emph{Critical}, \emph{High}, \emph{Medium}, and \emph{Low}. For each original report, we identify its security-severity level by extracting labels that start with \emph{Security\_Severity} from the \emph{AllLabels} field of the report. If a label in the format of \emph{Security\_Severity-}\emph{L} is available in the list of labels  (where \emph{L} is one of the four security-severity levels), then the severity of the report is \emph{L}. If no labels are available in the format of \emph{Security\_Severity-}\emph{L} in the list of labels, then we consider the severity of the report to be \emph{unclassified}. For each duplicate report~$D$, we use the security-severity level of the corresponding original report~$O$ instead of the security-severity level of the duplicate report~$D$. We exclude \emph{unclassified} vulnerabilities from the security-severity analysis. 

\paragraph{Firefox}

There are multiple security related keywords in the \emph{Keywords} field of a report. We only include reports in our analysis that contain one of the 6 following security tags in their \emph{Keywords} field: \emph{sec-critical, sec-high, sec-medium, sec-low, sec-vector, and sec-other}. 
If a report has one of the above keywords in its \emph{Keywords} field, we consider that report as a security report and include it for our analysis. There are some reports that have more than one security keyword. For those reports, we keep them as security reports in the dataset but exclude them from the analysis parts related to the security severity. 
We also realized that most of the collected reports that are opened in 2011 and before that year, do not have any security keywords assigned. Therefore, we only consider reports that are opened in 2012 and later for our analysis. For each duplicate report D, we use its corresponding original report's keyword as the duplicate security-severity field.
There are reports with other security related keywords in their \emph{Keywords} field which we exclude them since they are not actually security reports. For instance, reports which have \emph{sec-want} which is 'New features or improvement ideas related to security' according to Mozilla keywords explanation are excluded.\footnote{\url{https://bugzilla.mozilla.org/describekeywords.cgi}}

\subsubsection{Release Channels}
\label{sub_sec:others}

\paragraph{Chromium}

Google categorizes release versions as \emph{stable}, \emph{beta}, and \emph{dev}. Stable is the release that is available for end-users. Beta is the release that is available for a limited number of users to test features before releasing a stable release. Dev (commonly referred to as \emph{head}) is the release that is based on the last successful build. We use the term \emph{release channel} to refer to these release versions throughout the paper. Note that the term release version means the type of the release instead of the version number (e.g., Version 90 and Version 91).

Each security vulnerability \textit{I} impacts one or more release channels. To identify which security channel(s) is affected by vulnerability \textit{I}, we check labels in \emph{AllLabels} field that start with \emph{Security\_Impact}. Based on these, we identify three release channels during this process: stable, beta, and head. We group beta and head as development release channels and perform our analysis.

\paragraph{Firefox}
We followed the general approach we mentioned at the beginning of this section (\cref{generalapproach}). 

\subsubsection{Components}
\paragraph{Chromium}
For each report \textit{I}, we identify the components from its \emph{Component} field. Each report \textit{I} will contain a set of components $C_I$.  For each component \textit{c} in the $C_I$, we extract the set of the group, which indicates a list of all sub-levels from the top level to the bottom level of the component hierarchy. For example, if report I has a component \snippet{Internals$>$Plugins$>$PDF}, we extract the set of the group as \snippet{Internals}, \snippet{Internals$>$Plugins}, \snippet{Internals$>$Plugins$>$PDF}. We use $G_I$ to denote the set of all groups that correspond to all the components of the report \textit{I}. Some of the pairs of original report~$O$ and its duplicate report~$D$ have one of the following inconsistencies.

\begin{itemize}

\item If the \emph{Component} field of all duplicate reports of original report~$O$ are not the same and the \emph{Component} field of the original report~$O$ is empty. Still, all duplicate reports of the original report~$O$ contain the same set of groups of components $G_D$. We set the \emph{Component} field of the original report~$O$ to the value of the \emph{Component} field of duplicate reports.

\item If the \emph{Component} field of all duplicate reports of original report~$O$ are not the same and the \emph{Component} field of the original report~$O$ is not empty, then each pair of original report~$O$ and duplicate report~$D$, we check whether it satisfies at least one of the conditions. If it is satisfied, then we set the \emph{Component} field of duplicate report~$D$ to the value of the \emph{Component} field of original report~$O$ 

\begin{itemize}
    \item All the components of duplicate report~$D$ $(C_D)$ in the set $(C_O)$ or the set $(G_O)$.
    \item  All the groups of the components of duplicate report~$D$ $(G_D)$ present either in the set $(C_O)$ or the set $(G_O)$.
\end{itemize}

\end{itemize}

\paragraph{Firefox}
We followed the general approach we mentioned at the beginning of this section (\cref{generalapproach}).

\label{appendix_sub_sec:cleaning_timestamps}

\subsection{Earliest Report and Fixed Timestamps}
First reported Timestamp ($T_{earliest}$): the date and time of the first report of a vulnerability.
For each valid original report~$O$, we compute $T_{earliest}$ based on either one of the conditions:
\begin{itemize}
\item If the valid original report~$O$ reported before all of its duplicate reports, then we set $T_{earliest}$ to the value of \emph{OpenedTimestamp} field of the valid original report~$O$.
\item If the valid original report~$O$ is reported after one or more of its duplicates, we list all the timestamps from the \emph{OpenedTimestamp} field of all the duplicate reports of the valid original report~$O$, then set $T_{earliest}$ to the minimum timestamp from the list of all timestamps.
\end{itemize}
\paragraph{Fixed Timestamps In Firefox}
In the collected data, there are two vulnerabilities that both original and its duplicate has fixed time. We excluded those reports from our analysis related to fix. For some reports with \emph{WORKSFORME} status in Firefox, there are not specific fixed time. If that report has \emph{TrackingFlags} in its information page and it shows that the report got fixed in a version, we consider this report in our dataset as valid report, but we do not consider its fixed time.

\subsection{Rediscovery}
There are some vulnerabilities that are reported by the same origin reporter multiple times in both Chromium and Firefox. In the rediscovery analysis, we remove redundant reported reports and keep only the earliest report from the same origin.
There are also reports that do not have accessible pages in Bugzilla. Since for those reports we cannot identify which report is the earliest one, we excluded them from the rediscovery analysis.
In Firefox, some reports are incomplete with respect to replication and patching. For some of them, Mozilla opened a new report of the vulnerability, which was then completed with respect to this information, and marked the first report as a duplicate. We also exclude these vulnerabilities from our analysis since they are not actual rediscoveries.
\subsection{Exploited Vulnerabilities}

To identify exploited vulnerabilities, we first collect an initial set of exploited vulnerabilities from the Known Exploited Vulnerabilities Catalog of CISA\footnote{\url{https://www.cisa.gov/known-exploited-vulnerabilities-catalog}}. Then, we extend this set iteratively using a snowballing method by identifying terms in comments related to exploitation (e.g., \emph{exploited in the wild}) and gathering vulnerabilities whose comments include these terms. 
We manually verify the descriptions of these vulnerabilities to find false positives.
Finally, we restrict the set to valid original security vulnerabilities, resulting in a set of 18 and 37 exploited vulnerabilities for Firefox and Chromium respectively.

\fi

\end{document}